\documentclass[%
preprint,
amsmath,amssymb,  aps, prc, nofootinbib
]{revtex4-1}
\usepackage{graphicx}
\usepackage{dcolumn}
\usepackage{bm}

\usepackage{float}

\usepackage{xcolor}
\usepackage{hyperref}
\usepackage{tikz-feynman}
\tikzfeynmanset{compat=1.1.0}
\usepackage{tikz}
\usetikzlibrary{trees}
\usetikzlibrary{decorations.pathmorphing}
\usetikzlibrary{decorations.markings}
\definecolor{MyOrange}{HTML}{FF9900}
\definecolor{MyBlue}{HTML}{3366FF}
\definecolor{MyRed}{HTML}{CC0000}
\usepackage{subcaption}

\newcommand{\be}{\begin{equation}}
\newcommand{\ee}{\end{equation}}

\begin{document}

\title{Pseudoscalar and vector open charm mesons at finite temperature}
\author{Gl\`oria Monta\~na$^1$, \`Angels Ramos$^1$, Laura Tolos$^{2,3,4,5}$, and Juan M. Torres-Rincon$^2$}
\affiliation{$^1$Departament de F\'isica Qu\`antica i Astrof\'isica and Institut de Ci\`encies del Cosmos (ICCUB), Facultat de F\'isica,  Universitat de Barcelona, Mart\'i i Franqu\`es 1, 08028 Barcelona, Spain}
\affiliation{$^2$Institut f\"ur Theoretische Physik, Goethe Universit\"at Frankfurt, Max von Laue Strasse 1, 60438 Frankfurt, Germany}
\affiliation{$^3$Frankfurt Institute for Advanced Studies, Ruth-Moufang-Str. 1, 60438 Frankfurt am Main, Germany}
\affiliation{$^4$Institute of Space Sciences (ICE, CSIC), Campus UAB, Carrer de Can Magrans, 08193, Barcelona, Spain}
\affiliation{$^5$Institut d'Estudis Espacials de Catalunya (IEEC), 08034 Barcelona, Spain}
\date{\today}


\begin{abstract}
Vacuum and thermal properties of pseudoscalar and vector charm mesons are analyzed within a self-consistent many-body approach, employing a chiral effective field theory that incorporates heavy-quark spin symmetry. Upon unitarization of the vacuum interaction amplitudes for the scattering of charm mesons off light mesons in a fully coupled-channel basis, new dynamically generated states are searched. The imaginary-time formalism is employed to extend the calculation to finite temperatures up to $T=150$ MeV. Medium-modified spectral shapes of the $D$, $D^*$, $D_s$ and $D_s^*$ mesons are provided. The temperature dependence of the masses and decay widths of the nonstrange $D_0^*$ (2300) and $D_1^*$(2430) mesons,  both showing a double-pole structure in the complex-energy plane, is also reported, as well as that of the $D_{s0}^*$(2317) and $D_{s1}^*$(2460) resonances and other states not yet identified experimentally. Being the first calculation incorporating open-charm vector mesons at finite temperature in a self-consistent fashion, it brings up the opportunity to discuss the medium effects on the open charm sector under the perspective of chiral and heavy-quark spin symmetries.
\end{abstract}

\maketitle

\section{\label{intro}Introduction}

The in-medium properties of mesons with charm content have been a matter of high interest over the years (see  \cite{Rapp:2011zz,Tolos:2013gta,Hosaka:2016ypm,Aarts:2016hap} for reviews).  This interest was triggered because of the $J/\Psi$ suppression in heavy-ion collisions, as seen at Super Proton Synchrotron (SPS) energies by the NA50 Collaboration~\cite{Gonin:1996wn}, which was predicted in Ref.~\cite{Matsui:1986dk} as a signature of the existence of the quark-gluon plasma due to color screening. The $J/\Psi$ absorption in hot dense matter could be also modified due to the change of the properties of open-charm mesons in matter in the comover scattering scenario (see, for example, the initial works of Refs.~\cite{Capella:2000zp,Cassing:1999es,Vogt:1999cu,Gerschel:1998zi}), thus providing a complementary explanation for $J/\Psi$ suppression. 
Moreover, the possible attraction felt by $D$ mesons in nuclear matter could lead to the formation of  open-charm meson bound states in nuclei \cite{Tsushima:1998ru,GarciaRecio:2010vt,GarciaRecio:2011xt,Yamagata-Sekihara:2015ebw,Yasui:2012rw}. Also, the FAIR experimental facility in Germany with the CBM heavy-ion collision experiment will address the features of the phase QCD diagram using medium-modified open-charm mesons as a fundamental probe \cite{Rapp:2011zz}. The NA61 Collaboration at SPS energies, on the other hand, will also focus on open charm production in fixed-target heavy-ion collisions after upgrading their detector~\cite{Merzlaya:2019blz}.

Several theoretical works have addressed the properties of open-charm mesons in nuclear matter. Those works range from phenomenological estimates based on the quark-meson coupling model (see, for example, the initial works of~\cite{Guichon:1987jp,Sibirtsev:1999js}), nuclear mean-field calculations in matter \cite{Mishra:2003se,Kumar:2011ff,Pathak:2014vra}, Polyakov-loop extensions of Nambu-Jona-Lasinio models~\cite{Blaschke:2011yv}, models based on $\pi$-exchange implementing heavy-quark symmetries~\cite{Yasui:2012rw}, QCD sum-rule (QSR) computations (see~\cite{Gubler:2018ctz} for a recent review) to self-consistent unitarized coupled-channel approaches (for a review, see~\cite{Tolos:2013gta}).  The full spectral features of the open-charm mesons in nuclear matter emerge naturally from the latter ones. Starting from the exploratory works of Refs.~\cite{Tolos:2004yg,Tolos:2005ft}, the spectral features of open charm  in nuclear matter have been studied within $t$-channel vector-meson exchange unitarized models~\cite{Lutz:2005vx,Mizutani:2006vq,Tolos:2007vh,JimenezTejero:2011fc}. 

Later on, heavy-quark spin symmetry constraints were implemented explicitly in a unitarized coupled-channel approach~\cite{GarciaRecio:2008dp,Gamermann:2010zz,Romanets:2012hm,Garcia-Recio:2013gaa,GarciaRecio:2012db}.  Heavy-quark spin symmetry is a proper QCD spin-flavor symmetry that appears when the quark masses, such as the charm mass, become larger than the typical confinement scale. As a consequence of this symmetry, the spin interactions vanish for infinitely massive quarks. Thus, heavy hadrons come in doublets (if the spin of the light degrees of freedom is not zero), which are degenerated in the infinite quark-mass limit. The implementation of this symmetry led to the determination of open charm-nucleon interactions and open-charm spectral functions in nuclear matter in a consistent and reliable way
\cite{Tolos:2009nn,GarciaRecio:2010vt,GarciaRecio:2011xt}.

Other studies have been devoted to examine the properties of charmed mesons in a meson-rich environment. Most of these works, though, have been concentrated on the determination of the hidden-charm $J/\Psi$ dissociation cross sections in heavy-ion collisions (see~\cite{Rapp:2008tf} for a review). In fact, there are several studies on the $J/\Psi$-hadron interaction at finite temperature based on chiral Lagrangians~\cite{Haglin:2000ar,Bourque:2008es,Blaschke:2008mu}, quark model calculations~\cite{Zhou:2012vv,Maiani:2004py,Maiani:2004qj,Bourque:2008es} and QSR schemes~\cite{Duraes:2002px,Duraes:2002ux}, QCD lattice (see~\cite{Rothkopf:2019ipj} and references therein), and effective Lagrangians~\cite{Mitra:2014ipa,Abreu:2017cof}.  With regards to open-charm mesons, the studies on open-charm thermal relaxation in heavy-ion collisions~\cite{Laine:2011is,He:2011yi,Ghosh:2011bw,Abreu:2011ic,Tolos:2013kva,Ozvenchuk:2014rpa,Song:2015sfa} have triggered the interest of open charm at finite temperature. Investigations of open-charm mesons at finite temperature have been performed using QSR approaches \cite{Hilger:2011cq,Buchheim:2018kss,Gubler:2020hft} and calculations on the lattice~\cite{Bazavov:2014yba,Bazavov:2014cta,Kelly:2018hsi}. Also, effective models in a hot hadronic bath have been developed in Refs.~\cite{Mishra:2003se,Fuchs:2004fh,He:2011yi,Blaschke:2011yv,Ghosh:2013xea,Sasaki:2014asa}.

Recently, a finite-temperature unitarized approach based on a $SU(4)$ effective Lagrangian has been put forward \cite{Cleven:2017fun}, where the implications of pionic matter at finite temperatures on the properties of open and hidden charm mesons have been studied. Whereas the $J/\Psi$ stays narrow even at $T= 150$ MeV,  the  $D$ and $D^*$ mesons acquire a substantial width in the pionic bath, reaching 30--40 MeV at $T= 150$ MeV. 

In the present paper, we address the properties of open-charm mesons in a hot mesonic bath (mainly formed by pions), within a finite-temperature unitarized approach, following the path of Ref.~\cite{Cleven:2017fun} and extending our previous work in Ref.~\cite{Montana:2020lfi}. As compared to Ref.~\cite{Cleven:2017fun}, the dynamics of open charm with light mesons is based on an effective Lagrangian that is expanded up to next-to-leading order (NLO) in the chiral counting, while keeping  leading order (LO) in the heavy-quark expansion~\cite{Kolomeitsev:2003ac,Guo:2009ct,Geng:2010vw,Abreu:2011ic}. Moreover, the present paper extends our work of Ref.~\cite{Montana:2020lfi}, going beyond the analysis of $D_0^*(2300)$ and $D_s(2317)$ states and their possible identification as the chiral partners of the $D$ and $D_s$ ground states~\cite{Montana:2020lfi}. 

In this work we perform a detailed analysis of not only the scalar open-charm sector but all strange-isospin channels in the scalar and axial vector open-charm sectors in vacuum and at finite temperature by exploiting the heavy-quark spin symmetry. To this end, we use a unitarized approach based on a chiral effective field theory that implements heavy-quark spin symmetry at leading order. We start by analyzing the pole structure in vacuum of the dynamically-generated scalar excited states $D_0^*(2300)$ and $D_{s0}^* (2317)$ states as well as the axial vector excited ones $D_1^*(2430)$ and $D_{s1}^*(2460)$. We then extend our calculations to finite temperature by means of a self-consistent many-body approach and study the thermal dependence of the $D$, $D^*$, $D_s$ and $D_s^*$ mesons together with that of  $D_0^*(2300)$,  $D_{s0}^* (2317)$, $D_1^*(2430)$ and $D_{s1}^*(2460)$ states not only in a pionic bath, as in our previous work, but also in a pionic and kaonic medium.

The paper is organized as follows. In Sec.~\ref{formalism} we present the details of the effective Lagrangian and the unitarization procedure for open charm at zero and finite temperature, whereas in Sec.~\ref{results} we show our results for the scattering amplitudes and dynamically generated states at zero and finite temperature, paying a special attention to the spectral functions  of open-charm ground states as well as the finite-temperature mass and width modifications of open-charm ground states and dynamically generated ones. Finally, in Sec.~\ref{conclusions} we give our conclusions and future outlook.

\section{$D$-meson interaction with light mesons}
\label{formalism}

In this section we give details on the effective field theory used to describe the dynamics of $D$ and $D^*$ mesons, and their interactions with light mesons: $\pi,K,\bar{K}$ and $\eta$ mesons (we do not consider $\eta'$ mesons in this work because of their larger mass).

In the first part we present the effective Lagrangian involving the heavy and light mesons, which is based on chiral and heavy-quark symmetries and expanded up to NLO in the chiral counting and at LO in heavy-quark mass counting~\cite{Kolomeitsev:2003ac,Guo:2009ct,Geng:2010vw,Abreu:2011ic}. We describe how to fix the unknown low-energy constants from recent lattice-QCD calculations in the same sector~\cite{Guo:2018tjx}. Then, we provide the values of the tree-level scattering amplitudes for the different processes considered in this work. Finally, we describe how to construct unitarized amplitudes by solving a Bethe-Salpeter equation in a coupled-channel basis, at both $T=0$ and $T\neq0$. For the latter, we work within the imaginary-time formalism (ITF) requiring self-consistency of the heavy-meson self-energy.

We summarize in Table~\ref{tab:isospin_multiplets} the degrees of freedom (both heavy and light mesons) used in this work together with their vacuum masses.

\begin{table}[htbp!]
\begin{tabular}{|cccccc|cccccc|}
\hline
$J^P=0^-$ & Multiplet & \hspace{0.3cm}$I$\hspace{0.3cm} & \hspace{0.3cm}$S$\hspace{0.3cm} & \hspace{0.3cm}$C$\hspace{0.3cm} & $m\,\rm(MeV)$ & $J^P=1^-$ & Multiplet & \hspace{0.3cm}$I$\hspace{0.3cm} & \hspace{0.3cm}$S$\hspace{0.3cm} & \hspace{0.3cm}$C$\hspace{0.3cm} & $m\,\rm(MeV)$  \\
\hline
$\pi$ & $(\pi^+,\pi^0,\pi^-)$ & 1 & 0 & 0 & $138.04$ &
 &  &  &  &  &  \\
$\eta$ & $(\eta^0)$ & 0 & 0 & 0 & $547.86$ &
 &  &  &  &  &  \\
$K$ & $(K^{ +},K^{ 0})$ & $1/2$ & $+1$ & 0 & $495.64$ &
 &  &  &  &  &  \\
$\bar{K}$ & $(\bar{K}^{ 0},K^{ -})$ & $1/2$ & $-1$ & 0 & $495.64$ &
 &  &  &  &  &  \\
$D$ & $(D^{ +},D^{ 0})$ & $1/2$ & 0 & $+1$ & $1867.24$ & 
$D^*$ & $(D^{* +},D^{* 0})$ & $1/2$ & 0 & $+1$ & $2008.56$ \\
$\bar{D}$ & $(\bar{D}^{ 0},D^{ -})$ & $1/2$ & 0 & $-1$ & $1867.24$ &
$\bar{D}^*$ & $(\bar{D}^{* 0},D^{* -})$ & $1/2$ & 0 & $-1$ & $2008.56$ \\
$D_s$ & $(D_s^{+})$ & 0 & $+1$ & $+1$ & $1968.34$ &
$D_s^{*}$ & $(D_s^{*+})$ & 0 & $+1$ & $+1$ & $2112.20$ \\
$\bar{D}_s$ & $(D_s^{-})$ & 0 & $-1$ & $-1$ & $1968.34$ &
$\bar{D}_s^{*}$ & $(D_s^{*-})$ & 0 & $-1$ & $-1$ & $2112.20$ \\
\hline
\end{tabular}
\centering
\caption{Isospin multiplets of the light and heavy mesons considered in this work, together with the spin, parity, isospin, strangeness and charm quantum numbers; and the isospin-averaged values of the masses in the PDG book~\cite{Tanabashi:2018oca}. }
\label{tab:isospin_multiplets}
\end{table}

\subsection{Effective Lagrangian and tree-level scattering amplitudes}

Interactions between heavy and light mesons are described in terms of an effective Lagrangian for the  degrees of freedom described in Table~\ref{tab:isospin_multiplets}. It is based on chiral symmetry (involving the physics of the pseudo-Goldstone bosons at low energies) as well as heavy-quark spin-flavor symmetry for the charmed mesons (both pseudoscalar $D$ and vector $D^*$ mesons). In the heavy-quark-mass counting we only consider the LO effective Lagrangian, whereas in chiral power counting we will consider both LO and NLO orders. From now on, the terminology LO and NLO will exclusively refer to chiral power counting in the effective Lagrangian,
\be {\cal L}= {\cal L}_{\rm LO}+{\cal L}_{\rm NLO} \ . \ee

The LO Lagrangian contains the kinetic terms and interactions of the $D$ mesons as well as self-interactions of the pseudo-Goldstone bosons. For the latter, the pure light-meson sector is described by the standard chiral perturbation theory (ChPT)~\cite{Gasser:1983yg}, whose Lagrangian is not explicitly written here. The LO Lagrangian reads
\begin{align}
 \mathcal{L}_{\rm LO}&\ =\mathcal{L}^{\rm ChPT}_{\rm LO}+\langle\nabla^\mu D\nabla_\mu D^\dagger\rangle-m_D^2\langle DD^\dagger\rangle-\langle\nabla^\mu D^{*\nu}\nabla_\mu D^{*\dagger}_{\nu}\rangle+m_D^2\langle D^{*\nu}D^{*\dagger}_{\nu}\rangle \nonumber \\
 &\ +ig\langle D^{*\mu}u_\mu D^\dagger-Du^\mu D^{*\dagger}_\mu\rangle+\frac{g}{2m_D}\langle D^*_\mu u_\alpha\nabla_\beta D^{*\dagger}_\nu-\nabla_\beta D^*_\mu u_\alpha D^{*\dagger}_\nu\rangle\epsilon^{\mu\nu\alpha\beta} \ ,
\end{align}
where $D$ denotes the antitriplet of $0^-$ $D$-mesons, $D=\begin{pmatrix} D^0 & D^+ & D^+_s \end{pmatrix}$, and similarly for the vector $1^-$ states, $D^*_\mu=\begin{pmatrix} D^{*0} & D^{*+} & D^{*+}_s \end{pmatrix}_\mu$. The light mesons are encoded into $u_\mu=i(u^\dagger\partial_\mu u-u\partial_\mu u^\dagger)$, where $u=\exp (i\Phi/\sqrt{2} f_\pi)$ is the unitary matrix of pseudo-Goldstone bosons in the exponential representation,

\be \Phi = \left(
\begin{array}{ccc}
\sqrt{\frac12} \pi^0 +\sqrt{\frac16} \eta & \pi^+ & K^+ \\
\pi^- & -\sqrt{\frac12} \pi^0+\sqrt{\frac16} \eta & K^0 \\
K^- & \bar{K^0} & -\sqrt{\frac23} \eta \\
\end{array}
\right)\ , \ee
and $f_\pi$ is the pion decay constant, $f_\pi=92.4$ MeV. We note that in the matrix representation of the $SU(3)$ meson octet we have already identified the $\eta_8$ with the physical $\eta$ by neglecting the $\eta-\eta'$ mixing.

The angle brackets in the Lagrangian denote the trace in flavor space and the connection of the covariant derivative $\nabla_\mu D^{(*)}=\partial_\mu D^{(*)} -D^{(*)}\Gamma^\mu$ reads $\Gamma_\mu=\frac{1}{2}(u^\dagger\partial_\mu u+u\partial_\mu u^\dagger)$.

The NLO Lagrangian reads
\begin{align}\nonumber\label{eq:lagrangianNLO}
 \mathcal{L}_{\rm NLO}=&\ \mathcal{L}^{\rm ChPT}_{\rm NLO} -h_0\langle DD^\dagger\rangle\langle\chi_+\rangle+h_1\langle D\chi_+D^\dagger\rangle+h_2\langle DD^\dagger\rangle\langle u^\mu u_\mu\rangle \\ \nonumber
 &\ +h_3\langle Du^\mu u_\mu D^\dagger\rangle+h_4\langle\nabla_\mu D\nabla_\nu D^\dagger\rangle\langle u^\mu u^\nu\rangle+h_5\langle\nabla_\mu D\{u^\mu,u^\nu\}\nabla_\nu D^\dagger \rangle \\ \nonumber
 &\ +\tilde{h}_0\langle D^{*\mu}D^{*\dagger}_\mu\rangle\langle\chi_+\rangle-\tilde{h}_1\langle D^{*\mu}\chi_+D^{*\dagger}_\mu\rangle-\tilde{h}_2\langle D^{*\mu}D^{*\dagger}_\mu\rangle\langle u^\nu u_\nu\rangle \\ 
 &\ -\tilde{h}_3\langle D^{*\mu}u^\nu u_\nu D^{*\dagger}_\mu\rangle-\tilde{h}_4\langle\nabla_\mu D^{*\alpha}\nabla_\nu D^{*\dagger}_\alpha\rangle\langle u^\mu u^\nu\rangle-\tilde{h}_5\langle\nabla_\mu D^{*\alpha}\{u^\mu,u^\nu\}\nabla_\nu D^{*\dagger}_\alpha\rangle,
\end{align}
where ${\mathcal L}^{\rm ChPT}_{\rm NLO}$ represents the NLO ChPT Lagrangian involving only $\Phi$, and $\chi_+=u^\dagger\chi u^\dagger+u\chi u$, with the quark-mass matrix $\chi={\rm diag}(m_\pi^2,m_\pi^2,2m_K^2-m_\pi^2)$. For more details we recommend Refs.~\cite{Guo:2009ct,Geng:2010vw,Abreu:2011ic,Liu:2012zya,Tolos:2013kva}.

The tree-level amplitudes are extracted from the LO+NLO Lagrangian and they will be kept at strictly lowest-order in heavy-quark mass expansion, i.e. we will only consider amplitudes ${\cal O}(1/m_D^{0},1/m_{D^*}^{0})$.  They have already been given in the cited references. For convenience we reproduce them here once more. Since we are working with the lowest-order amplitudes, there are no tree level diagrams converting $D$ mesons into $D^*$~\cite{Abreu:2011ic}, and the two sectors are independent, but related by heavy-quark spin symmetry. For a binary scattering involving a charm meson with incoming channel $i$ and outgoing channel $j$ the amplitudes read
\begin{align} \nonumber\label{eq:potential}
 V^{ij}(s,t,u)=&\ \frac{1}{f_\pi^2}\Big[\frac{C_{\rm LO}^{ij}}{4}(s-u)-4C_0^{ij}h_0+2C_1^{ij}h_1\\ 
 &\ -2C_{24}^{ij}\Big(2h_2(p_2\cdot p_4)+h_4\big((p_1\cdot p_2)(p_3\cdot p_4)+(p_1\cdot p_4)(p_2\cdot p_3)\big)\Big)\\ \nonumber
 &\ +2C_{35}^{ij}\Big(h_3(p_2\cdot p_4)+h_5\big((p_1\cdot p_2)(p_3\cdot p_4)+(p_1\cdot p_4)(p_2\cdot p_3)\big)\Big)
 \Big],
\end{align}
where $s=(p_1+p_2)^2,t=(p_1-p_3)^2,u=(p_1-p_4)^2$ are the Mandelstam variables. The different channels $i,j$ we consider in this work are summarized in Table~\ref{tab:channels}. The isospin coefficients $C^{ij}$ are given in Table~\ref{tab:coeffiso}, whereas in Appendix~\ref{app:isospincoeff} we provide the isospin coefficients in the charge basis. 
  
\begin{table}[htbp!]
 \begin{tabular}{|c|cc|cc|}
 \hline
$(S,I)$ & Channels ($J^P=0^-$) & Threshold (MeV) & Channels ($J^P=1^-$) & Threshold (MeV) \\
\hline
$(-1,0)$  & $D\bar{K}$ & 2364.88 & $D^* \bar{K}$ & $2504.20$ \\
$(-1,1)$  & $D\bar{K}$ & 2364.88 & $D^* \bar{K}$ & $2504.20$ \\
$(0,\frac12)$ & $D\pi$ & 2005.28 & $D^* \pi$ & $2146.59$ \\
        & $D\eta$ & 2415.10 & $D^* \eta$ & $2556.42$ \\
        & $D_s\bar{K}$ & 2463.98 & $D_s^* \bar{K}$ & $2607.84$ \\
$(0,\frac32)$ & $D\pi$ & 2005.28 & $D^* \pi$ & $2146.59$ \\
$(1,0)$   & $DK$ & 2364.88 & $D^*K$ & $2504.20$ \\
        & $D_s\eta$ & 2516.20 & $D_s^* \eta$ & $2660.06$ \\
$(1,1)$   & $D_s\pi$ & 2106.38 & $D_s^* \pi$ & $2250.24$ \\
        & $DK$ & 2364.88 & $D^* K$ & $2504.20$ \\
$(2,\frac12)$ & $D_sK$ & 2463.98 & $D_s^* K$ & $2607.84$ \\
\hline
 \end{tabular}
 \centering
 \caption{Meson-meson channels considered in this work together with the threshold energy, their total spin, strangeness, and isospin quantum numbers. We provide the channels involving the pseudoscalar ($J^P=0^-$) $D$ meson and the vector ($J^P=1^-$) $D^*$ meson.}
 \label{tab:channels}
 \end{table}
 
\begin{table}[htbp!]
\begin{tabular}{|c l c c c c c|}
 \hline
 $(S,I)$  &  Channel $i\rightarrow j$ & \hspace{0.3cm}$C_{\rm LO}^{ij}$\hspace{0.3cm} & \hspace{0.3cm}$C_0^{ij}$\hspace{0.3cm} & \hspace{0.3cm}$C_1^{ij}$\hspace{0.3cm} & \hspace{0.3cm}$C_{24}^{ij}$\hspace{0.3cm} &  \hspace{0.3cm}$C_{35}^{ij}$\hspace{0.3cm}  \\  
\hline
 $(-1,0)$  & $D\bar{K}\rightarrow D\bar{K}$ & $-1$ & $m_K^2$ & $m_K^2$ & $1$ & $-1$ \\
 $(-1,1)$ & $D\bar{K}\rightarrow D\bar{K}$ & $1$ & $m_K^2$ & $-m_K^2$ & $1$ & $1$ \\
 $(0,\frac{1}{2})$  & $D\pi\rightarrow D\pi$ & $-2$ & $m_\pi^2$ & $-m_\pi^2$ & $1$ & $1$ \\
   & $D\pi\rightarrow D\eta$ & $0$ & $0$ & $-m_\pi^2$ & $0$ & $1$ \\
   & $D\pi\rightarrow D_s\bar{K}$ & $-\sqrt{\frac{3}{2}}$ & $0$ & $-\frac{\sqrt{3}}{2\sqrt{2}}(m_K^2+m_\pi^2)$ & $0$ & $\sqrt{\frac{3}{2}}$ \\
   & $D\eta\rightarrow D\eta$ & $0$ & $m_\eta^2$ & $-\frac{1}{3}m_\pi^2$ & $1$ & $\frac{1}{3}$ \\
   & $D\eta\rightarrow D_s\bar{K}$ &$-\sqrt{\frac{3}{2}}$ & $0$ & $\frac{1}{2\sqrt{6}}(5m_K^2-3m_\pi^2)$ & $0$ & $-\frac{1}{\sqrt{6}}$ \\
   & $D_s\bar{K}\rightarrow D_s\bar{K}$ & $-1$ & $m_K^2$ & $-m_K^2$ & $1$ & $1$ \\
 $(0,\frac{3}{2})$  & $D\pi\rightarrow D\pi$ & $1$ & $m_\pi^2$ & $-m_\pi^2$ & $1$ & $1$ \\
 $(1,0)$   & $DK\rightarrow DK$ & $-2$ & $m_K^2$ & $-2m_K^2$ & $1$ & $2$ \\
   & $DK\rightarrow D_s\eta$ & $-\sqrt{3}$ & $0$ & $-\frac{1}{2\sqrt{3}}(5m_K^2-3m_\pi^2)$ & $0$ & $\frac{1}{\sqrt{3}}$ \\
   & $D_s\eta\rightarrow D_s\eta$ & $0$ & $m_\eta^2$ & $-\frac{4}{3}(2m_K^2-m_\pi^2)$ & $1$ & $\frac{4}{3}$ \\
 $(1,1)$  & $D_s\pi\rightarrow D_s\pi$ & $0$ & $m_\pi^2$ & $0$ & $1$ & $0$ \\
   & $D_s\pi\rightarrow DK$ & $1$ & $0$ & $-\frac{1}{2}(m_K^2+m_\pi^2)$ & $0$ & $1$ \\
   & $DK\rightarrow DK$ & $0$ & $m_K^2$ & $0$ & $1$ & $0$ \\
 $(2,\frac{1}{2})$  & $D_sK\rightarrow D_sK$ & $1$ & $m_K^2$ & $-m_K^2$ & $1$ & $1$ \\
\hline
\end{tabular}
\caption{Isospin coefficients $C_{ij}$ for the sectors with strangeness $S$ and isospin $I$.}
\label{tab:coeffiso}
\end{table}

The low-energy constants (LECs) $h_i$ with $i=0,...,5$ have been revisited in this work in view of the recent studies~\cite{Guo:2009ct,Liu:2012zya,Guo:2018tjx} based on lattice-QCD data. The LEC $h_0$ is determined in the latest two references through the fit of lattice-QCD data for the masses of the $D$ and $D_s$ at different pion masses, while the value of $h_1$ can be fixed from the physical mass splitting between the $D$ and $D_s$, $h_1=(m_{D_s}^2-m_D^2)/[4(m_K^2-m_\pi^2)]$. Linear combinations of the remaining LECs, $h_{24}=h_2+h_4/\bar{M}_D^2$ and $h_{35}=h_3+2h_5/\bar{M}_D^2$, with $\bar{M}_D^2=(M_D+M_{D_s})/2$, are determined in~\cite{Guo:2009ct,Liu:2012zya} by fits to the scattering lengths calculated on the lattice, simultaneously also to lattice finite-volume energy levels in~\cite{Guo:2018tjx}. We have taken the values of the LECs from the Fit-2B in~\cite{Guo:2018tjx} for which the full amount of lattice data available and the physical value of $f_\pi$ are considered, and which is the preferred fit of the authors according to large $N_c$ arguments. The values are shown in Table~\ref{tab:LECs} for the sake of completeness, for both pseudoscalar and vector charmed mesons, where we have considered $h_i=\tilde{h}_i$ at LO in the heavy-quark expansion but used the different physical values of $\bar{M}_D$ and $\bar{M}_{D^*}$ in the determination of $h_4$, $h_5$, $\tilde{h}_4$ and $\tilde{h}_5$. The difference between our unitarized amplitudes and those in~\cite{Guo:2018tjx} lies in the regularization procedure, as will be explained in the next section.

\begin{table}[htbp!]
\begin{tabular}{|l c c c c c c|}
 \hline
 & \hspace{0.3cm}$h_0$\hspace{0.3cm} & \hspace{0.3cm}$h_1$\hspace{0.3cm} & \hspace{0.3cm}$h_2$\hspace{0.3cm} & \hspace{0.3cm}$h_3$\hspace{0.3cm} & \hspace{0.3cm}$h_4\rm\,(MeV^{-2})$\hspace{0.3cm} & \hspace{0.3cm}$h_5\rm\,(MeV^{-2})$\hspace{0.3cm} \\
\hline
$D\Phi$ & $0.033$ & $0.45$ & $-0.12$ & $1.67$ & $-0.0054\cdot10^{-6}$ & $-0.22\cdot10^{-6}$ \\
$D^*\Phi$ &  $0.033$ & $0.45$ & $-0.12$ & $1.67$ & $-0.0047\cdot10^{-6}$ & $-0.19\cdot10^{-6}$ \\ 
\hline
\end{tabular}
\caption{Values of the LECs used in this work (from Fit-2B in~\cite{Guo:2018tjx}) for the interaction of pseudoscalar (first row) and vector (second row) charmed mesons with the light mesons.}
\label{tab:LECs}
\end{table}

To conclude this section, we comment on the extraction of the $S$-wave component of these tree-level scattering amplitudes. The $S$-wave projected amplitude is computed as
\be V^{ij}(s)= \frac{1}{2} \int_{-1}^{1} d(\cos \theta) \ V^{ij}(s,t(s,\cos \theta)) P_{L=0} (\cos \theta) \ , \label{eq:proj} \ee
where $t(s,\cos \theta)$ is given in terms of $s$, $\cos \theta$ and the meson masses cf.~\cite{Guo:2018tjx}, $u=\sum_i m_i^2-s-t$ and $P_L (\cos \theta)$ is the Legendre polynomial of order $L$ normalized to $\int_{-1}^1  dx P_L (x) P_{L'} (x)=2\delta_{LL'}/(2L+1)$. In particular, for the $S$-wave projection, we only need $P_{L=0}=1$. 

The analysis of higher partial waves is left for future studies, as in the present work we focus on dynamically generated resonances that can correspond to some observed states listed in the PDG \cite{Tanabashi:2018oca}. In the energy region explored here, no potential candidates are found that can correspond to states generated dynamically from the meson-meson interaction in partial waves higher than $L=0$. In addition, we note that, for a particular channel, the P-wave interaction kernel is of reduced strength compared to the S-wave one, especially close to threshold, around which the molecular states appear.   
  
\subsection{Unitarization and self-consistent propagators}

Let us motivate the need of a unitarization technique for the scattering amplitudes. The amplitudes found in Eq.~(\ref{eq:potential}) can describe the meson-meson scattering at low energies but they do not satisfy the exact unitarity condition. For the $s$-wave projected amplitude (Eq.~(\ref{eq:proj})) this condition would read (for the one-channel case $D\Phi$) $\textrm{Im } V(s) = \rho_{D\Phi}(s) |V(s)|^2$, where $\rho_{D\Phi}(s)$ is the two-body phase space. This relation is evidently not satisfied because $V(s)$ are real. This causes an unphysical increase of the cross sections at intermediate energies. In addition, the polynomial structure of $V^{ij}(s)$ cannot generate resonances, which are signaled by the presence of poles (singularities) in the scattering amplitude. A unitarization method aims at constructing from $V^{ij}(s)$ a new amplitude $T^{ij}(s)$ which satisfies (for the 1-channel case $D\Phi$)
\be \textrm{Im } T(s) = \rho_{D\Phi}(s) \ |T(s)|^2 \ . \ee
This is a standard method used by the so-called unitarized chiral perturbation theories with great success since many years~\cite{Oller:2000ma}. Among the different unitarization methods, we employ the one based on the Bethe-Salpeter equation because, within a straightforward extension of field theory, it can be simply applied to finite temperature. Upon unitarization, $T^{ij}(s)$ helps to extend the limit of validity in energy because of the increase of the amplitude is tamed, and the cross sections saturate. Being a rational function of $s$, the $T$-matrix also allows for the dynamical generation of resonances.

We summarize here the unitarization procedure which allows us to fulfill exact unitarity in all scattering amplitudes. Starting from the tree-level amplitudes $V^{ij}(s)$ we construct the unitarized ones $T^{ij}(s)$ by solving a Bethe-Salpeter equation for the two-body problem in a full coupled-channel basis. We distinguish the cases at $T=0$ and $T\neq0$ as their methodologies and ulterior analyses differ considerably.

\subsubsection{Unitarization in vacuum}

In the on-shell factorization approach~\cite{Oller:1997ti,Oset:1997it}, the Bethe-Salpeter equation for the unitarized amplitude at $T=0$ reads:
\be \label{eq:BetheSalpeter} T^{ij} (s)= V^{ij} (s) + V^{il}(s) G^l(s) T^{lj} (s) \ , \ee
where $i,j$ represent the incoming and outgoing channels (see Table~\ref{tab:channels}), and we sum over the possible intermediate channels $l$. The two-body propagator function in vacuum is the loop function
\be\label{eq:loopVac}
  G^l(s)=i\int\frac{d^4q}{(2\pi)^4} \frac{1}{q^2-m_D^2+i\epsilon} \frac{1}{(p-q)^2-m_\Phi^2+i\epsilon} \ ,
\ee
with $p^\mu=(E,\vec{p})$. We make explicit that at $T=0$ the loop function is given as a function of the Mandelstam variable $s=p^2$. This function should be regularized in vacuum, for which one can employ a hard momentum cutoff $\Lambda$, and the corresponding expression of the loop reads
\begin{align}\label{eq:loopVacCut}\nonumber
  G^l_{\Lambda}(s)&\ = \frac{1}{32\pi^2}\Bigg\{-\frac{m_\Phi^2-m_D^2}{s}\ln\frac{m_D^2}{m_\Phi^2}+\ln\frac{m_D^2m_\Phi^2}{\Lambda^4} \\ \nonumber
  &\ +\frac{\sigma}{s}\bigg[\ln\big(s+(m_\Phi^2-m_D^2)+\sigma\sqrt{1+m_\Phi^2\Lambda^{-2}}\big)+\ln\big(s-(m_\Phi^2-m_D^2)+\sigma\sqrt{1+m_D^2\Lambda^{-2}}\big) \\ \nonumber
  &\ -\ln \big(-s-(m_\Phi^2-m_D^2)+\sigma\sqrt{1+m_\Phi^2\Lambda^{-2}}\big)-\ln\big(-s+(m_\Phi^2-m_D^2)+\sigma\sqrt{1+m_D^2\Lambda^{-2}}\big)\bigg] \\ 
  &\ +2\frac{(m_\Phi^2-m_D^2)}{s}\ln\frac{1+\sqrt{1+m_D^2\Lambda^{-2}}}{1+\sqrt{1+m_\Phi^2\Lambda^{-2}}} -2\ln\bigg[\bigg(1+\sqrt{1+\frac{m_D^2}{\Lambda^2}}\bigg)\bigg(1+\sqrt{1+\frac{m_\Phi^2}{\Lambda^2}}\bigg)\bigg]\Bigg\} \ ,
\end{align}
with $\sigma=\sqrt{\big(s-(m_D+m_\Phi)^2\big)\big(s-(m_D-m_\Phi)^2\big)}$. Alternatively, the loop integral can be calculated in dimensional regularization (DR):
 \begin{align}\label{eq:loopVacDR} \nonumber
  G^l_{\rm DR}(s; a_l(\mu))=&\ \frac{1}{16\pi^2} \Big\{ a_l(\mu)+\ln\frac{m_D^2}{\mu^2}+\frac{m_\Phi^2-m_D^2+s}{2s}\ln\frac{m_\Phi^2}{m_D^2}+ \\ \nonumber
   &\ +\frac{q_l}{\sqrt{s}}\left[\ln \big(s+(m_\Phi^2-m_D^2)+2q_l\sqrt{s}\big)+\ln \big(s-(m_\Phi^2-m_D^2)+2q_l\sqrt{s}\big)-\right. \\
   &\ \left. -\ln \big(-s-(m_\Phi^2-m_D^2)+2q_l\sqrt{s}\big)-\ln \big(-s+(m_\Phi^2-m_D^2)+2q_l\sqrt{s}\big) \right] \Big\} \ ,
 \end{align}
where $a_l(\mu)$ is the channel-dependent subtraction constant at the regularization scale $\mu$ for channel $l$  and $q_l=\sigma/(2\sqrt{s})$ is the on-shell three-momentum of the meson in the loop. If one demands that both regularization procedures give the same value of the loop function at threshold, the following expression for the subtraction constants is obtained:
\be\label{eq:subtconstant} a_l(\mu)=16\pi^2 [G^l_\Lambda(s^l_{\rm thr})-G^l_{\rm DR}(s^l_{\rm thr}; a_l=0)] \ , \ee 
for a given $\mu$. Notice that the running of $a_l(\mu)$ cancels the explicit $\mu$ dependence in Eq.~(\ref{eq:loopVacDR}), so the loop function does not depend on the regularization scale.

In Ref.~\cite{Guo:2018tjx}, the loop function is regularized with DR and the subtraction constants are considered as fit parameters together with the LECs. In a different approach, here we use the cutoff regularization scheme so as to follow the same convention as for $T > 0$, where the loop function is usually regularized by limiting the integrals to $|\vec{q}\,|< \Lambda$. We adjust the cutoff value to a representative scale of the degrees of freedom that are implicitly integrated out in the construction of the meson-meson interaction amplitude from the effective Lagrangian. Indeed, the pointlike interaction of Eq.~(\ref{eq:potential}) could have also been obtained from a $t$-channel diagram, involving two cubic meson vertices and the propagator of a vector meson of mass $m_V\sim m_\rho$, in the limit $m_V^2\gg t $ with $t$ being the four-momentum exchanged in the process. With this rationale in mind, our cutoff value is taken to be of the order of the $\rho$ meson mass, namely $\Lambda=800$~MeV, for all the channels. We have checked that the equivalent values of the subtraction constants obtained using Eq.~(\ref{eq:loopVacDR}) are compatible with those employed in~\cite{Guo:2018tjx}, with $\mu= 1$~GeV, and that the reproduction of scattering observables with our prescription is of comparable quality. 
This agreement reinforces the choice of $\Lambda=800$~MeV as an appropriate selection, a value which, in turn, determines the position of the dynamically generated resonances. 

In the on-shell approximation~\cite{Oller:1997ti} the unitarized scattering amplitude has an algebraic solution
\be\label{eq:tmatrixT0} T_{ij}(s) = V_{ik}(s) [1-V(s) G(s)]^{-1}_{kj} \ , \ee
which, in the general coupled-channel case, is a matrix equation.

Notice that the internal propagators of the loop function should include self-energy corrections due to interactions in vacuum. This effect dresses (and renormalizes) the mass of the mesons. At $T=0$ the dressed masses are simply the physical ones given in Table~\ref{tab:isospin_multiplets}, and there is no need to perform a self-consistent procedure. This is different in the $T\neq 0$ case.

The unitarization process leads to the potential emergence of poles in the resummed amplitude $T_{ij}(s)$ at the zeros of the denominator of Eq.~(\ref{eq:tmatrixT0}). These poles correspond to states that are dynamically generated by the attractive coupled-channel meson-meson interactions. The characterization of these states requires to analytically continue the $T$-matrix to the complex energy plane, where the search for poles should be performed in the correct Riemann sheet (RS). The loop function in Eq.~(\ref{eq:loopVac}) is a multivalued function with two RSs above threshold. To select a particular RS one needs to add a contribution to the imaginary part~\cite{Roca:2005nm},
\begin{align}\nonumber
G^l_{\rm II}(\sqrt{s}+i\varepsilon)=G^l_{\rm I}(\sqrt{s}-i\varepsilon)=[G^l_{\rm I}(\sqrt{s}+i\varepsilon)]^*=&\ G^l_{\rm I}(\sqrt{s}+i\varepsilon)-i2\,{\rm Im}\,G^l_{\rm I}(\sqrt{s}+i\varepsilon) \\
=&\ G^l_{\rm I}(\sqrt{s}+i\varepsilon)+i\frac{q_l}{4\pi\sqrt{2}} \ ,
\end{align}
where the subindices I and II denote the first and second RSs, respectively. The same result follows from changing the sign of the momentum $q_l$ in Eq.~(\ref{eq:loopVacDR}) or $\sigma$ in Eq.~(\ref{eq:loopVacCut}) and taking the phase prescription of the logarithms $\ln z=\ln|z|+i\theta$ as $0\le\theta<2\pi$.

The analytic structure of $G^l$ provides the unitarized amplitude $T_{ij}(s)$ with a set of $2^n$ RSs, where $n$ is the number of coupled channels. 
We define the {\it second RS} of the $T(s)$ amplitude as the one which is connected to the real energy axis from below and is obtained by using $G^l_{\rm I}$ for ${\rm Re } \sqrt{s}< m_{\Phi_l} + M_{D_l}$ and $G^l_{\rm II}$ for ${\rm Re } \sqrt{s}> m_{\Phi_l} + M_{D_l}$. 
This prescription provides the pole position and half-width closer to the values obtained from a Breit-Wigner parametrization of the associated amplitude \cite{Roca:2005nm}. However, not too far from a threshold channel to which a resonance couples strongly, the pole might appear in an RS  for which the transition from $G^l_{\rm I}$ to $G^l_{\rm II}$ is not applied and yet the resonance has a visible effect in the real-axis amplitudes. This is the case for the second, higher energy poles associated to the $D_0^*(2300)$ and the $D_1^*(2430)$, as will be discussed in Sec.~\ref{results}.

In the complex $\sqrt{s}\equiv \sqrt{z}$ plane, the real and imaginary parts of the pole positions $\sqrt{z_p}$, give the mass and the half-width of the dynamically generated states, respectively:
\be M_R={\rm Re\,}\sqrt{z_p} \ ,\quad \Gamma_R/2={\rm Im\,} \sqrt{z_p} \ .\ee
The poles located on the real axis of the physical RS, below the lowest threshold, correspond to {\it bound states}, those on an unphysical sheet at a necessarily complex energy correspond to {\it resonances}, and {\it virtual states} are poles that lie on the real axis of an unphysical sheet, below the lowest threshold. Resonance poles that are located on the unphysical sheet closest to the physical sheet (the second RS) are the ones that, together with bound states, are more likely to generate structures in the scattering amplitude. Therefore it is common to call resonances only the resonant poles in the second sheet and generalize the term virtual state to resonant poles in any other unphysical sheet, which can still yield to structure and cusps near the thresholds.

The scattering amplitude can be expanded in a Laurent series around the pole position
\be
T^{ij}(z)=\frac{g_ig_j}{z-z_p}+\sum_{n=0}^\infty T_{ij}^{(n)}(z-z_p)^n \ ,
\ee
where $g_i$ is the coupling of the resonance or bound state to the channel $i$ and $g_ig_j$ is the residue around the pole. Therefore, from the residue of the different components of the $T$-matrix around the pole one can extract the coupling constants to each of the channels,
\be
g_i^2=\Big[\frac{\partial}{\partial z^2}\Big(\frac{1}{T_{ii}(z)}\Big)\Big|_{z_p}\Big]^{-1} \ .
\ee

The concept of compositeness of shallow bound states was formulated by Weinberg in Ref.~\cite{Weinberg:1962hj}, applied to narrow resonances close to threshold in~\cite{Baru:2003qq,Hanhart:2011jz}, and subsequently extended to the complex pole position of a resonance by analytical continuation in~\cite{Hyodo:2011qc,Aceti:2012dd,Aceti:2014ala,Sekihara:2014kya}. An appropriate unitary transformation~\cite{Guo:2015daa} permits assigning real values to the compositeness of complex poles lying in the second Riemann sheet as
\be
\chi_i= \left|g_i^2\frac{\partial G_i(z_p)}{\partial z}\right| \ ,
\label{eq:compo}
\ee
which effectively measures the amount of $i{\rm th}$ channel meson-meson component in the dynamically generated state.

\subsubsection{Finite-temperature case}

For $T\neq0$ we need to account for several modifications, both in the methodology and in the final analysis of the dynamically generated states. We use the ITF, where the time dimension is Wick rotated and compactified in the range $[0,\beta=1/T]$, where $T$ is the temperature of the system. Any integration over the energy variable is therefore transformed into a summation over the so-called Matsubara frequencies, $q^0\rightarrow i\omega_n=2n\pi T i$ for bosons (see~\cite{Weldon:1983jn,Das:1997gg,kapustagale,lebellac,fetterwalecka} for details of the formalism). 

The tree-level amplitudes remain unmodified, but the two-body loop function in the Bethe-Salpeter equation is now modified as compared to the vacuum case. It reads
\be\label{eq:loopT}
  G_{D\Phi}(i\omega_m,\vec{p};T)=-T\sum_n \int\frac{d^3q}{(2\pi)^3} \frac{1}{\omega_n^2+{\vec{q}\,}^2+m_D^2} \frac{1}{(\omega_m-\omega_n)^2 +(\vec{p}-\vec{q}\,)^2+m_\Phi^2} \ .
\ee

Before performing the Matsubara summation over $\omega_n$ we introduce the Lehmann representation for the propagators in terms of the spectral function,
\be \label{eq:propagatorLehman}
  \mathcal{D}_M(i\omega_n,\vec{q}\,)=\int d\omega'\frac{S_M(\omega',\vec{q}\,)}{i\omega_n-\omega'}=\int_0^\infty d\omega'\frac{S_M(\omega',\vec{q}\,)}{i\omega_n-\omega'}-\int_0^\infty d\omega'\frac{S_{\bar{M
}}(\omega',\vec{q}\,)}{i\omega_n+\omega'} \ ,
\ee
where the subindex $M$ denotes the meson ($D$ or $\Phi$) and in the second equality we have separated the particle and antiparticle parts. Using delta-type spectral functions, $S_M(\omega,\vec{q}\,)=\frac{\omega_M}{\omega}\delta(\omega^2-\omega_M^2)$, with $\omega_M=\sqrt{\vec{q}\,^2+m_M^2}$, it is straightforward to see that Eq.~(\ref{eq:loopT}) is recovered. By keeping generic spectral functions $S_D(\omega,\vec{q}\,)$ and $S_\Phi(\omega',\vec{p}-\vec{q}\,)$ the Matsubara summation using Cauchy's residue theorem gives the following expression for the loop function
\begin{align} \nonumber
 G_{D\Phi}(i\omega_m,\vec{p};T)&=\int\frac{d^3q}{(2\pi)^3}\int_0^\infty d\omega\int_0^\infty d\omega'\Bigg\{\frac{S_D(\omega,\vec{q}\,)S_{\Phi}(\omega',\vec{p}-\vec{q}\,)}{i\omega_m-\omega-\omega'}\big[1+f(\omega,T)+f(\omega',T)\big] \\ \nonumber
 & +\frac{S_{\bar{D}}(\omega,\vec{q}\,)S_{\Phi}}{i\omega_m+\omega-\omega'}\big[f(\omega,T)-f(\omega',T)\big] 
   -\frac{S_{D}(\omega,\vec{q}\,)S_{\bar{\Phi}}(\omega',\vec{p}-\vec{q}\,)}{i\omega_m-\omega+\omega'}\big[f(\omega,T)-f(\omega',T)\big] \\
 & -\frac{S_{\bar{D}}(\omega,\vec{q}\,)S_{\bar{\Phi}}(\omega',\vec{p}-\vec{q}\,)}{i\omega_m+\omega+\omega'}\big[1+f(\omega,T)+f(\omega',T)\big] \Bigg\} \ ,
\end{align}
where $f(\omega,T)=1/\left(\exp ( \omega/T ) -1\right)$ is the equilibrium Bose-Einstein distribution function.

We use $f(-\omega,T)=-[1+f(\omega,T)]$ and $S_{\bar M}(-\omega,\vec{q};T)=-S_M(\omega,\vec{q};T)$ and analytically continue the external Matsubara frequency to real energies $i\omega_m \rightarrow E + i\epsilon$ so as to write the expression above in the following compact way:
\begin{align} \label{eq:loopT2}
  G_{D\Phi}(E,\vec{p};T)&=\int\frac{d^3q}{(2\pi)^3}\int d\omega\int d\omega'\frac{S_{D}(\omega,\vec{q};T)S_{\Phi}(\omega',\vec{p}-\vec{q};T)}{E-\omega-\omega'+i\varepsilon} [1+f(\omega,T)+f(\omega',T)] \ ,
\end{align}
where the integrals over energy extend from $-\infty$ to $+\infty$.

At finite temperature the meson masses are dressed by the medium. The effects of finite temperature in the unitarized scattering amplitudes are readily obtained by solving Eq.~(\ref{eq:BetheSalpeter}) with finite-temperature loops containing dressed mesons. Notice that in the ITF, as the thermal corrections enter in loop diagrams~\cite{lebellac,kapustagale}, the tree-level scattering amplitudes remain the same as in vacuum (with the zeroth component of the four-momentum expressed as a bosonic Matsubara frequency).

The spectral function of the heavy meson is computed from the imaginary part of its retarded propagator,
\begin{equation} \label{eq:specfunc}
  S_{D}(\omega,\vec{q};T)=-\frac{1}{\pi}{\rm Im\,}\mathcal{D}_{D}(\omega,\vec{q};T)=-\frac{1}{\pi}{\rm Im\,}\Bigg(\frac{1}{\omega^2-\vec{q}\,^2-m_{D}^2-\Pi_{D}(\omega,\vec{q};T)}\Bigg) \ ,
\end{equation}
where the heavy-meson self-energy follows from closing the light-meson line in the $T$-matrix.

In this work, we are interested in analyzing the medium modification of the $D$-meson propagator due to light mesons. However light mesons also suffer medium modifications and their spectral functions are modified by the interactions among themselves. For a pion gas we can use previous results in the literature to convince that the mass modification is small so that the use of a free spectral function is justified. In Appendix~\ref{app:modpion} we show this fact, where up to temperatures of $T=150$ MeV the pion mass varies at most 10 $\%$. At the top $T=150$ MeV we have used $m_\pi=120$ MeV mass in our numerical calculation and found that the final charm meson masses (both ground and the dynamically generated states) are modified less than 0.1 $\%$. Therefore, for this work we find good enough to use the free pion spectral function. A more thorough study (include mass modification of all Goldstone bosons as well as their thermal widths) will be performed in the future. 

For temperatures below $T_c$ the largest contribution to the $D$-meson dressing will come from pions, as the abundance of heavier light mesons, i.e. kaons and eta mesons, is suppressed by the Bose factors. We note that the contribution of a kaonic bath can be relevant for temperatures close to $T_c$.  In order to study this contribution, in Sec.~\ref{results} we will analyze  the modification of open-charm mesons in the presence of a kaonic bath by taking into account the corresponding Bose-Einstein distribution.
 
The pion contribution to the heavy-meson self-energy is computed in the ITF:
\begin{equation}
 \Pi_D(i\omega_n,\vec{q};T)=-T\int\frac{d^3q'}{(2\pi)^3}\sum_m\mathcal{D}_\pi(\omega_m-\omega_n,\vec{q}\,')T_{D\pi}(\omega_m,\vec{p}\,) \ ,
\end{equation}
where $\vec{q}\,'=\vec{p}-\vec{q}$ is the three-momentum of the pion.
It is convenient to use the Lehman representation for the pion propagator introduced in Eq.~(\ref{eq:propagatorLehman}) as well as for the T-matrix,
\begin{equation}
 T_{D\pi}(i\omega_m,\vec{p}\,)=-\frac{1}{\pi}\int dE \ \frac{{\rm Im\,} T_{D\pi}(E,\vec{p}\,)}{i\omega_m-E} \ ,
\end{equation}
and, following the same procedure as for the loop function described above, the expression obtained for the heavy-meson self-energy reads
\begin{align}\label{eq:selfE_4terms}\nonumber
    \Pi_{D}(i\omega_n,\vec{q};T)&= \frac{1}{\pi} \int\frac{d^3q'}{(2\pi)^3}\frac{1}{2\omega_\pi}\int_0^\infty dE
    \Bigg\{\frac{1+f(E,T)+f(\omega_\pi,T)}{E-i\omega_n+\omega_\pi}
    +\frac{1+f(E,T)+f(\omega_\pi,T)}{E+i\omega_n+\omega_\pi} \\ 
    &-\frac{f(E,T)-f(\omega_\pi,T)}{E-i\omega_n-\omega_\pi}
    -\frac{f(E,T)-f(\omega_\pi,T)}{E+i\omega_n-\omega_\pi}\Bigg\}
    {\rm Im\,}T_{D\pi}(E,\vec{p};T) \ ,
\end{align}
with $\omega_\pi = \sqrt{q'^2+m_\pi^2}$. This expression can be condensed in the following,
\begin{equation}\label{eq:selfE}
    \Pi_{D}(\omega,\vec{q};T)=\frac{1}{\pi} \int\frac{d^3q'}{(2\pi)^3}\int dE\frac{\omega}{\omega_\pi}\frac{f(E,T)-f(\omega_\pi,T)}{\omega^2-(\omega_\pi-E)^2+i\varepsilon \textrm{sgn} (\omega)} \ {\rm Im\,}T_{D\pi}(E,\vec{p};T) \ ,
\end{equation}
after the analytical continuation $i\omega_n\rightarrow \omega+i\varepsilon$.  Note that we only consider the Bose-Einstein distribution for pions at finite temperature while neglecting other possible medium modifications. 

The self-energy needs to be regularized as it also contains the vacuum contribution. In order to do so, we separate the vacuum and matter parts. The vacuum contribution is identified with the expression surviving in the limit $T\rightarrow 0$. One can see from Eq.~(\ref{eq:selfE_4terms}) that in this limit the Bose factors in the statistical weights of the numerators cancel, and only the terms with a factor 1 remain. Our prescription for the regularization is to take these terms to zero at finite temperature, as they were effectively included in the renormalization of the $D$-meson mass at $T=0$.

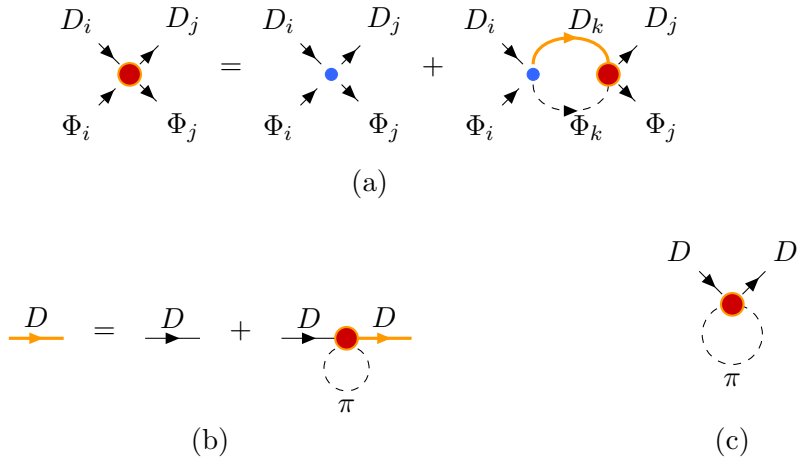
\begin{figure}[htbp!]
\centering 
\begin{subfigure}[b]{\textwidth}\centering 
\captionsetup{skip=0pt}
 \begin{tikzpicture}[baseline=(i.base)]
    \begin{feynman}[small]
      \vertex (a) {\(D_i\)};
      \vertex [below right = of a] (i) {};
      \vertex [above right = of i] (b) {\(D_j\)};
      \vertex [below right = of i] (d) {\(\Phi_j\)};
      \vertex [below left=of i] (c) {\(\Phi_i\)};
      \diagram* {
        (a) -- [fermion] (i), 
        (i) -- [fermion] (b),
        (c) -- [charged scalar] (i),
        (i) -- [charged scalar] (d),
       };
     \draw[dot,minimum size=4mm,thick,MyOrange,fill=MyRed] (i) circle(1.5mm);
    \end{feynman}
  \end{tikzpicture}
  $=$
  \begin{tikzpicture}[baseline=(i.base)]
    \begin{feynman}[small]
      \vertex (a) {\(D_i\)};
      \vertex [below right = of a] (i) {};
      \vertex [above right = of i] (b) {\(D_j\)};
      \vertex [below right = of i] (d) {\(\Phi_j\)};
      \vertex [below left=of i] (c) {\(\Phi_i\)};
      \diagram*{
        (a) -- [fermion] (i), 
        (i) -- [fermion] (b),
        (c) -- [charged scalar] (i),
        (i) -- [charged scalar] (d),
       } ;    
     \draw[dot,MyBlue,fill=MyBlue] (i) circle(.8mm);
    \end{feynman}
  \end{tikzpicture}
  $+$
  \begin{tikzpicture}[baseline=(i.base)]
    \begin{feynman}[small]
      \vertex (a) {\(D_i\)};
      \vertex [below right = of a] (i) {};
      \vertex [right = of i] (j) {};
      \vertex [above right = of j] (b) {\(D_j\)};
      \vertex [below right = of j] (d) {\(\Phi_j\)};
      \vertex [below left=of i] (c) {\(\Phi_i\)};
      \vertex [above right = of i] (b1) {\(D_k\)};
      \vertex [below right=of i] (d1) {\(\Phi_k\)};
      \diagram*{
        (a) -- [fermion] (i), 
        (j) -- [fermion] (b),
        (i) -- [MyOrange, fermion, very thick, half left, looseness=1.2] (j),
        (i) -- [charged scalar, half right, looseness=1.2] (j),
        (c) -- [charged scalar] (i),
        (j) -- [charged scalar] (d),
       } ;    
     \draw[dot,MyBlue,fill=MyBlue] (i) circle(0.8mm);
     \draw[dot,minimum size=4mm,thick,MyOrange,fill=MyRed] (j) circle(1.5mm);
    \end{feynman}
  \end{tikzpicture}
\caption{}
\label{fig:subfigA}
\end{subfigure}
\\[0.2cm]
\hspace{-1cm}\begin{subfigure}[b]{0.7\textwidth}\centering 
\captionsetup{skip=-5pt}
 \begin{tikzpicture}[baseline=(a.base)]
    \begin{feynman}[small]
      \vertex (a) {};
      \vertex [right = of a] (b) {};
      \diagram* {
        (a) -- [MyOrange, fermion, very thick,edge label=\(\textcolor{black}{D}\)] (b), 
       };
    \end{feynman}
  \end{tikzpicture}
  $=$
 \begin{tikzpicture}[baseline=(a.base)]
    \begin{feynman}[small]
      \vertex (a) {};
      \vertex [right = of a] (b) {};
      \diagram* {
        (a) -- [fermion, edge label=\(D\)] (b), 
       };
    \end{feynman}
  \end{tikzpicture}
  $+$
  \begin{tikzpicture}[baseline=(a)]
    \begin{feynman}[small, inline=(a)]
      \vertex (a) {};
      \vertex [right = of a] (i) {};
      \vertex [right = of i] (b) {};
      \vertex [below = 0.4cm of i] (d) {};
      \vertex [below = 0.9cm of i] (e) {\(\pi\)};
      \diagram* {
        (a) -- [fermion,edge label=\(D\)] (i), 
        (i) -- [MyOrange, fermion, very thick,edge label=\(\textcolor{black}{D}\)] (b),
       } ;   
     \draw[dashed] (d) circle(0.3cm);
     \draw[dot,minimum size=4mm,thick,MyOrange,fill=MyRed] (i) circle(1.5mm);
    \end{feynman}
  \end{tikzpicture}
\caption{}
\label{fig:subfigB}
\end{subfigure}\hspace{-1cm}
\begin{subfigure}[b]{0.25\textwidth}\centering
\captionsetup{skip=5pt}
  \begin{tikzpicture}
    \begin{feynman}[small]
      \vertex (a) {\(D\)};
      \vertex [below right = of a] (i) {};
      \vertex [above right = of i] (b) {\(D\)};
      \vertex [below = 0.4cm of i] (d) {};
      \vertex [below = 1cm of i] (e) {\(\pi\)};
      \diagram* {
        (a) -- [fermion] (i), 
        (i) -- [fermion] (b),
       } ;   
     \draw[dashed] (d) circle(0.4cm);
     \draw[dot,minimum size=4mm,thick,MyOrange,fill=MyRed] (i) circle(1.5mm);
    \end{feynman}
  \end{tikzpicture}
\caption{}
\label{fig:subfigC}
\end{subfigure}
\caption{(a) Bethe-Salpeter equation. The $T$-matrix is solved self-consistently with dressed internal heavy-meson propagator.  (b) Dressed heavy-meson propagator. (c) Heavy-meson self-energy. The heavy meson is dressed by the unitarized interaction with pions ($T$-matrix, red dot).}
\end{figure}

This set of equations is solved iteratively until self-consistency is obtained. The procedure is sketched in Figures~\ref{fig:subfigA}, \ref{fig:subfigB} and \ref{fig:subfigC}. The $T$-matrix amplitude is represented by a red blob, whereas the perturbative amplitude $V(s)$ is denoted with a blue dot. Figure~\ref{fig:subfigA} shows the Bethe-Salpeter equation for the two-body scattering. The intermediate propagator of the $D$-meson (yellow solid line) is itself dressed by interactions as shown in Fig.~\ref{fig:subfigB}, where the $T$-matrix is used in the Dyson equation for the propagator, giving rise to a self-consistent set of equations. As illustrated in Fig.~\ref{fig:subfigC} only the pion contribution, as being the dominant excitation in the thermal bath, is considered in the $D$-meson self-energy.

\section{Results}
\label{results}

We start this section by presenting our results for the scattering amplitudes and dynamically generated states at $T=0$ for both $J=0$ and $J=1$ charmed sectors. We then show the spectral functions of pseudoscalar and vector open-charm ground-state mesons at finite temperature. We finally determine the evolution with temperature of the masses and widths of not only the ground-state mesons but also the dynamically generated states.


\subsection{Scattering amplitudes and dynamically generated states at $T=0$}

In this section we focus on the $T=0$ case and distinguish the scalar $D$ and vector $D^*$ cases separately. Notice that these two sectors are not mixed when keeping the scattering diagrams at lowest order in the heavy-mass expansion, as argued before Eq.~({\ref{eq:potential}).

\subsubsection{$J=0$ case: Interactions and $D^*(2300)$ and $D_{s0}^*(2317)$ }

We start by analyzing the $D$ and $D_s$ interaction with  noncharmed pseudoscalar mesons for two different sectors given by total strangeness $S=0$ (Fig.~\ref{fig:VGT_0_05}), corresponding to the $D \pi$, $D \eta$ and $D_s \bar K$ coupled-channels calculation, and $S=1$ (Fig.~\ref{fig:VGT_1_0}), built from the $D K$ and $D_s \eta$ channels. We focus on those two sectors since we obtain several resonant states, among them two that can be identified with the experimental $D_0^*(2300)$ and $D_{s0}^*(2317)$, as we will discuss later on.

\begin{figure}[htbp!]
\includegraphics[width=0.6\textwidth]{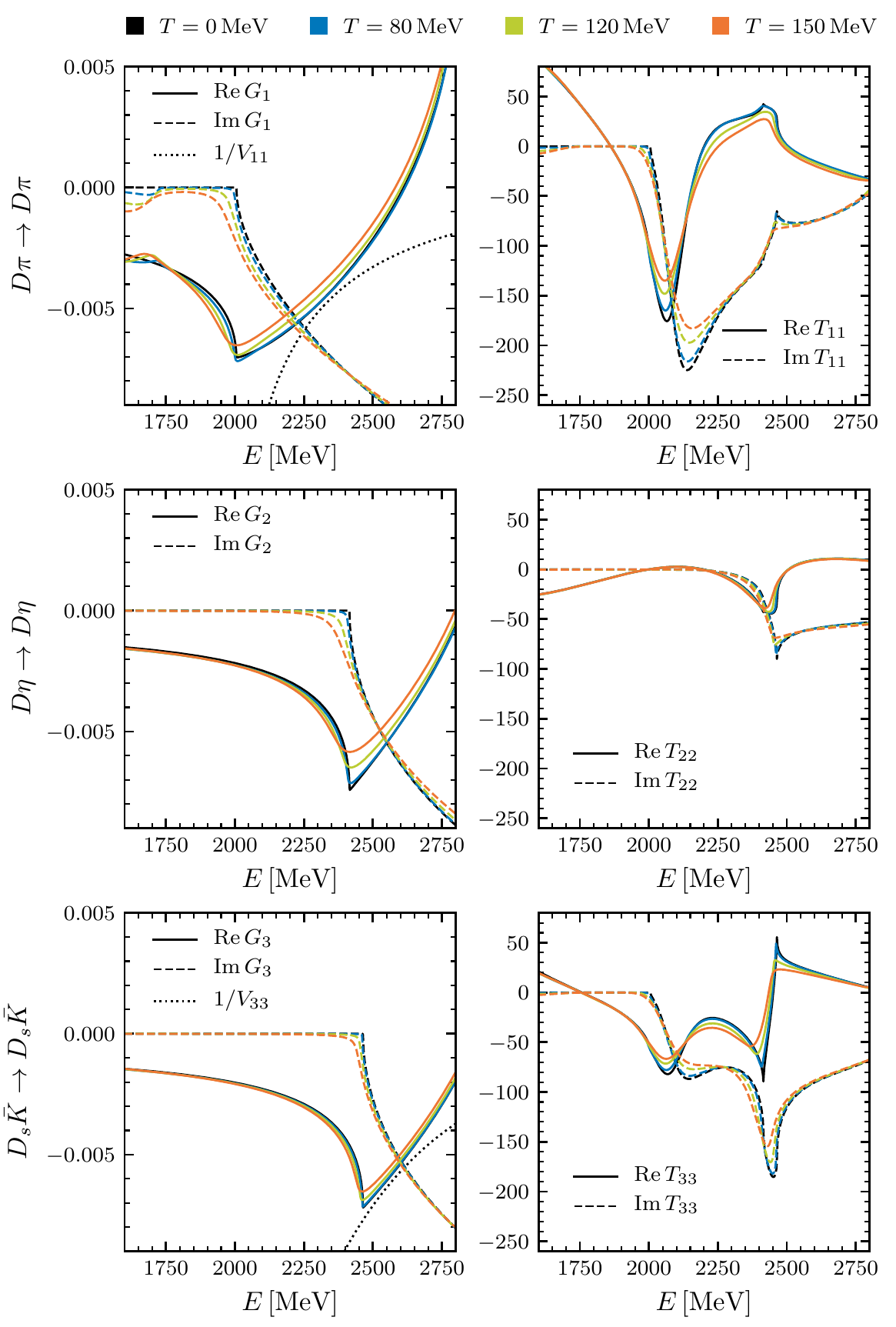}
\caption{The inverse of the interaction kernel, $V_{ii}$, the real and imaginary parts of the loop function, $G_i$, and the real and imaginary parts of the diagonal components of the T-matrix, $T_{ii}$, in the sector with strangeness and isospin $(S,I)=(0,\frac12)$, at various temperatures. The subindices 1, 2, 3 refer to the channels $D\pi$, $D\eta$ and $D_s\bar{K}$, respectively.}
\label{fig:VGT_0_05}
\end{figure}

The diagonal amplitudes in the strangeness $S=0$ and isospin $I=1/2$ sector are represented in the panels on the right-hand side (RHS) of Fig~\ref{fig:VGT_0_05} as functions of the total energy and for a center-of-mass momentum $\vec{P}=0$ and various temperatures. In order to understand the structures appearing in these amplitudes it is convenient to analyze first the energy dependence of the loop function $G$, which is displayed in the panels on the left-hand side (LHS) of the same figure. The imaginary part of the loop function (dashed lines) starts to have a significant strength from the value of $m_D+m_\Phi$ onward, which is the energy at which the right-hand $T=0$ unitarity cut starts. We note that, at finite temperature, a left-hand cut starting at $m_D-m_\Phi$ opens up. For the energy range displayed in Fig.~\ref{fig:VGT_0_05}, this cut can only be seen in the top panel, corresponding to the $D\pi$ channel, and it is more visible as the temperature increases. 

The inverse of the diagonal interaction kernel is also displayed in the panels on the LHS (dotted line) if it falls within the vertical scale employed for each channel.  In an uncoupled channel calculation, one should expect an enhancement in the corresponding unitarized amplitude when this quantity equals or becomes very close to the real part of the loop function (solid lines). This is just the reflection on the real energy axis of the pole generated by the solution of Eq.~(\ref{eq:tmatrixT0}). The consideration of coupled channels, apart from modifying slightly the energy position of the structures, makes them to be present in all the amplitudes, with more or less intensity depending on the coupling strength of the pole to each particular channel.  As can be seen in the RHS panels, in this $(S,I)=(0,\frac{1}{2})$ sector, besides cusps related to thresholds that are especially visible in the real part of the amplitudes (solid lines), we clearly see two enhancements  in the imaginary part (dashed lines), at around 2100 and 2450 MeV, which are connected to poles of the amplitude in the complex plane, as we will discuss below.

   \begin{figure}[htbp!]
   \includegraphics[width=0.6\textwidth]{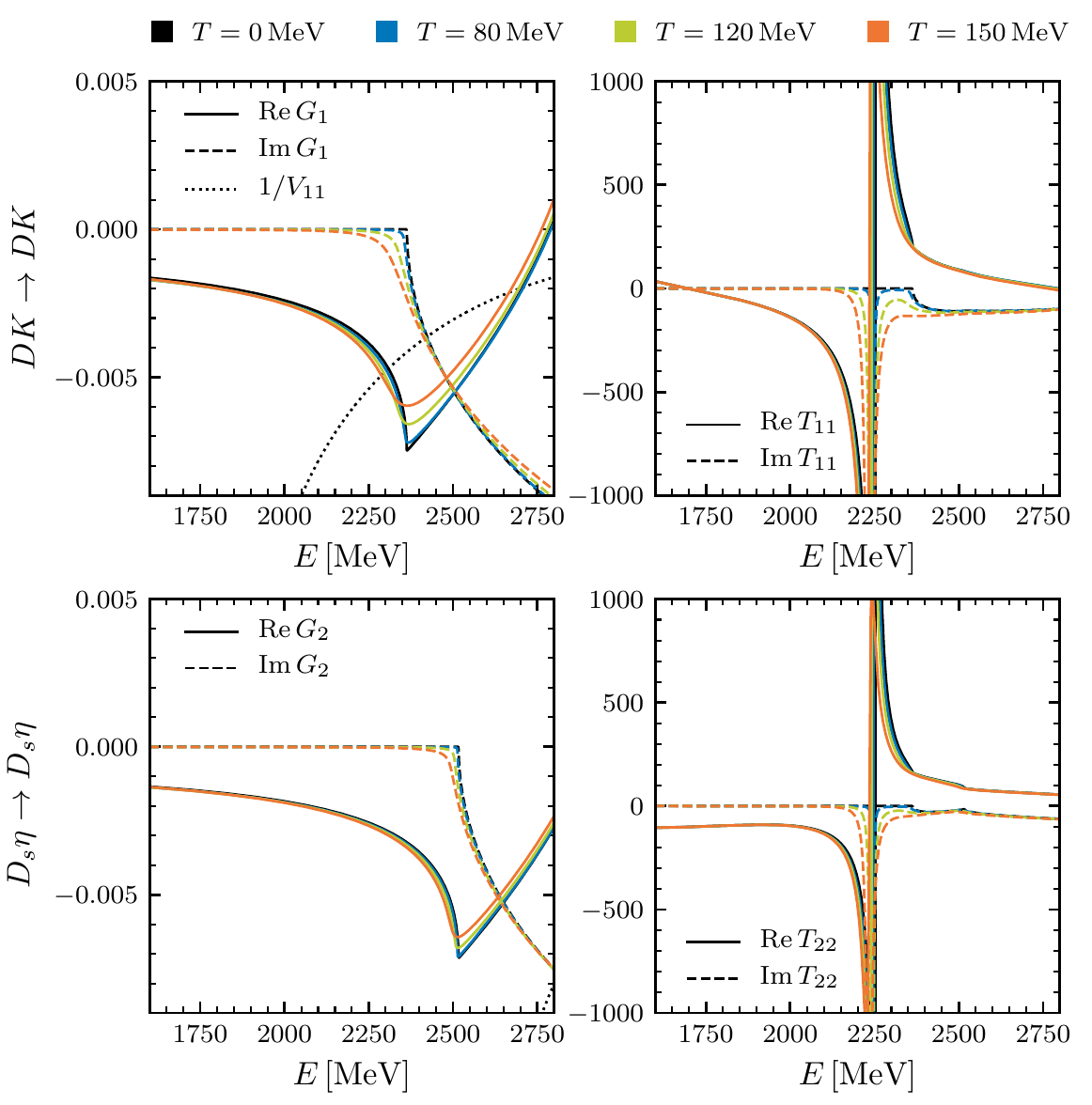}
   \caption{The same as in Fig.~\ref{fig:VGT_0_05} in the sector with $(S,I)=(1,0)$. The subindices 1, 2 refer to the channels $D_s\pi$ and $DK$, respectively.}
   \label{fig:VGT_1_0}
   \end{figure}

Our results for the strangeness $S=1$ and isospin $I=0$ sector are shown in Fig.~\ref{fig:VGT_1_0}. In this case we only observe a narrow structure around 2250 MeV, tied to the position of the crossing of the inverse of the $D K$ potential with the real part of the loop function, which occurs below the $m_D+m_K$ threshold and therefore leads to a bound sate in the real axis at $T=0$, as we will see.

\begin{table}[htbp!]
\begin{tabular}{ccccccc}
\hline
& $(S,I)$ & RS & $M_R$ & $\Gamma_R/2$ & $|g_i|$ & $\chi_i$ \\
& & & (MeV) & (MeV) & (GeV) &  \\
\hline
\hline
 & $(-1,0)$ & $(-)$ & $2261.5$ & $102.9$ & $|g_{D\bar{K}}|=11.6$ & $\chi_{D\bar{K}}=0.43$ \\
\hline
$D_0^*(2300)$ & $(0,\frac12)$ & $(-,+,+)$ & $2081.9$ & $86.0$  & $|g_{D\pi}|=8.9$  & $\chi_{D\pi}=0.40$ \\
 &    &           &        &       & $|g_{D\eta}|=0.4$  & $\chi_{D\eta}=0.00$ \\
 &   &           &        &       & $|g_{D_s\bar{K}}|=5.4$  & $\chi_{D_s\bar{K}}=0.05$ \\
 &   & $(-,-,+)$ & $2529.3$ & $145.4$ & $|g_{D\pi}|=6.7$  & $\chi_{D\pi}=0.10$ \\
 &   &           &        &       & $|g_{D\eta}|=9.9$  & $\chi_{D\eta}=0.40$ \\
 &   &           &        &       & $|g_{D_s\bar{K}}|=19.4$ & $\chi_{D_s\bar{K}}=1.63$ \\
    \hline
$D_{s0}^*(2317)$ &   $(1,0)$ & $(+,+)$ & $2252.5$ & $0.0$ & $|g_{DK}|=13.3$ & $\chi_{DK}=0.66$ \\
 &    &           &        &       & $|g_{D_s\eta}|=9.2$ & $\chi_{D_s\eta}=0.17$ \\
\hline
 & $(1,1)$ & $(-,+)$ & $2264.6$ & $200.9$ & $|g_{D_s\pi}|=7.3$ & $\chi_{D_s\pi}=0.21$ \\
 &         &         &          &         & $|g_{DK}|=5.9$     & $\chi_{DK}=0.08$ \\
\hline
 \end{tabular}
 \centering
 \caption{Dynamically generated poles in the $J=0$ sector. We find 5 poles in the different $(S,I)$ channels. The first column is reserved for those poles that can be associated to a state listed in the PDG~\cite{Tanabashi:2018oca}. RS denotes the Riemann sheet of the pole with the convention given in the main text. $M_R$ and $\Gamma_R/2$ are the real and imaginary parts of the pole location in the complex energy plane, $g_i$ denotes the effective coupling to different channels and $\chi_i$ is the compositeness of the state. }
 \label{tab:poles0}
 \end{table}

As discussed in the previous plots, apart from threshold effects, the different structures that appear in the scattering amplitudes correspond to the presence of poles or dynamically generated states that appear due to the attractive coupled-channel meson-meson interactions. 
The five poles that we find in the $J=0$ sector at $T=0$ are summarized in Table~\ref{tab:poles0}. The first column of this table indicates the possible experimental assignment of the poles according to the PDG ~\cite{Tanabashi:2018oca}, whereas the second column shows the strangeness and isospin content of the state. The third column indicates the RS where the pole is found, with the convention that the RS of the loop function for each of the coupled channels is indicated as $+$ for first and $-$ for second. In the fourth and fifth columns we give the mass $M_R$ and width $\Gamma_R/2$ of the state, while in the fifth column  $g_i$ denotes the effective coupling to different channels and  in the sixth column $\chi_i$ is the compositeness of the state.

In $(S,I)=(-1,0)$ sector, we find one virtual state, as the pole lies below the lowest threshold but it appears in the unphysical (-) RS. 
In the $(S,I)=(0,\frac12)$ sector we find two poles that correspond to the $D_0^* (2300)$ state. This double pole structure of the $D_0^* (2300)$ is well documented~\cite{Tanabashi:2018oca}, being our results compatible with those given in Refs.~\cite{Guo:2018tjx,Albaladejo:2016lbb}. For the position of the lower pole, we find that the real part lies between the $D\pi$ and $D\eta$ thresholds and
it has a sizable imaginary part, which is a consequence of the large value of the coupling of the generated resonance to the $D\pi$ channel, to which it can decay.
The mass of the higher pole is above the last threshold, that is, the $D_s \bar{K}$ one, and also has a large decay width, as it couples sizably to the channels opened for its decay. However, this pole appears in the $(-,-,+)$ RS, with this RS being only connected to the real axis between the $D\eta$ and the $D_s\bar{K}$ thresholds. In fact, for different values of the parameters~\cite{Guo:2018tjx,Albaladejo:2016lbb} this pole appears between the $D\eta$ and $D_s\bar{K}$ thresholds or even below the $D\eta$ threshold. Moreover, it is worth noticing that
the lower pole qualifies mainly as a $D\pi$ state, as indicated by the large value of the compositeness, whereas the higher one is essentially $D_s\bar{K}$ system, although we should note that this case does not correspond to a canonical resonance in the sense that the associated pole does not reside in the RS that is directly accessible from the physical one. Therefore, as discussed in Ref.~\cite{Guo:2015daa},  the physical interpretation of Eq.~(\ref{eq:compo}) as a probabilistic compositeness  is not valid for this resonance, a fact that is corroborated by the value larger than one obtained in this case.
In the $(S,I)=(1,0)$ sector we find only one pole, which lies on the real axis below the $DK$ threshold, that is in the $(+,+)$ RS. It is identified with the $D_{s0}^*(2317)$ resonance, and has sizable couplings to both $DK$ and $D_s \eta$, as given by the compositeness. With the present model, the pole mass turns out to be smaller than that of the experimental resonance, but small variation of the parameters can easily accommodate this state to the observed position, in line with similar models in the literature that have advocated this resonance to be mostly a $DK$ hadronic molecule (see Ref.~\cite{Guo:2017jvc} and references therein).
In the $(S,I)=(1,1)$ sector, there is a resonance in the (-,+) RS with a large width, as it couples strongly to $D_s \pi$ states to which it can decay. This resonance cannot be identified with any of the PDG states known up to now.

  \begin{figure}[htbp!]
   \includegraphics[width=0.6\textwidth]{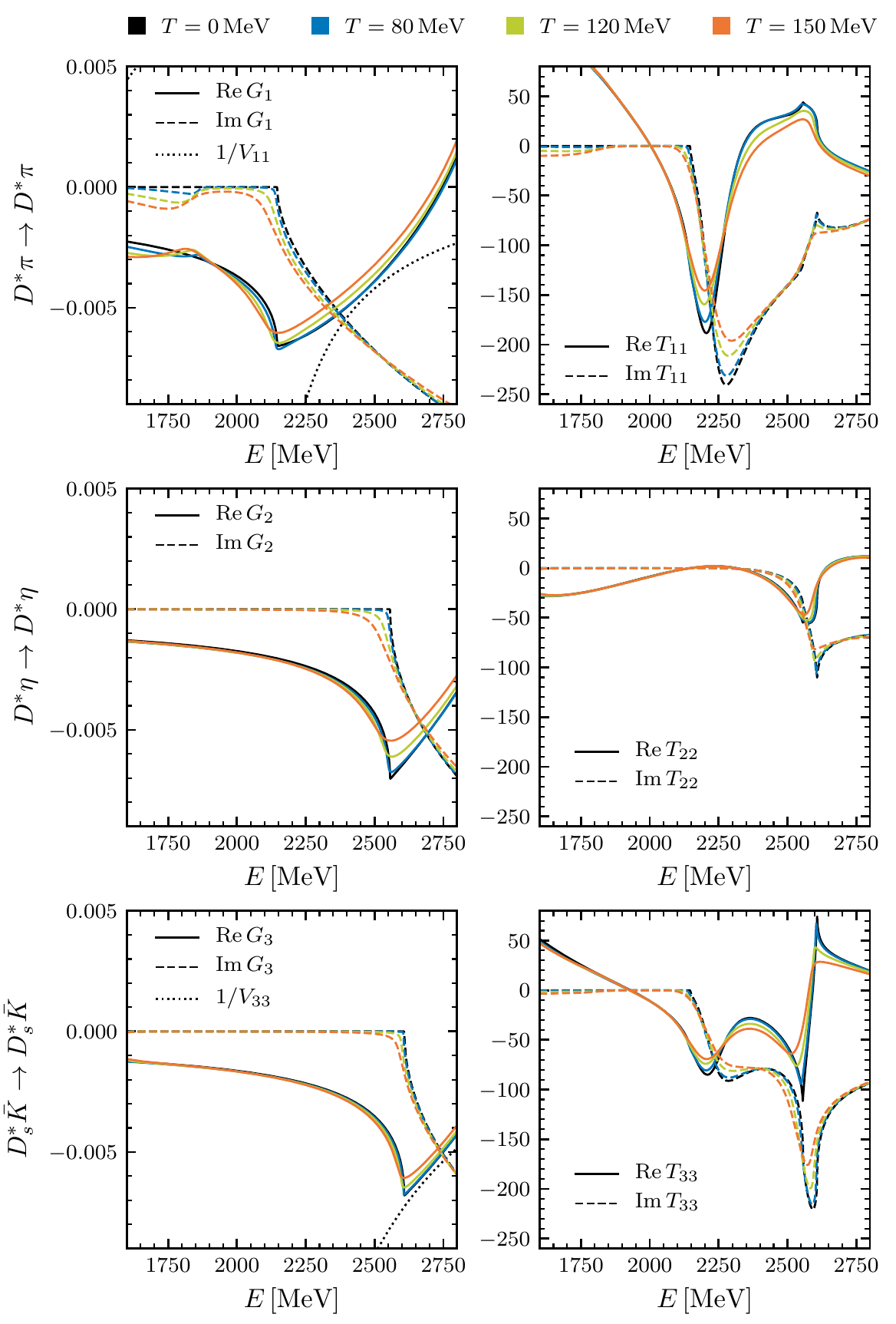}
   \caption{The inverse of the interaction kernel, $V_{ii}$, the real and imaginary parts of the loop function, $G_i$, and the real and imaginary parts of the diagonal components of the T-matrix, $T_{ii}$, in the sector with strangeness and isospin $(S,I)=(0,\frac12)$, at various temperatures. The subindices 1, 2, 3 refer to the channels $D^*\pi$, $D^*\eta$ and $D_s^*\bar{K}$, respectively.}
   \label{fig:VGT_0_05_vec}
   \end{figure}
 
     \begin{figure}[htbp!]
   \includegraphics[width=0.6\textwidth]{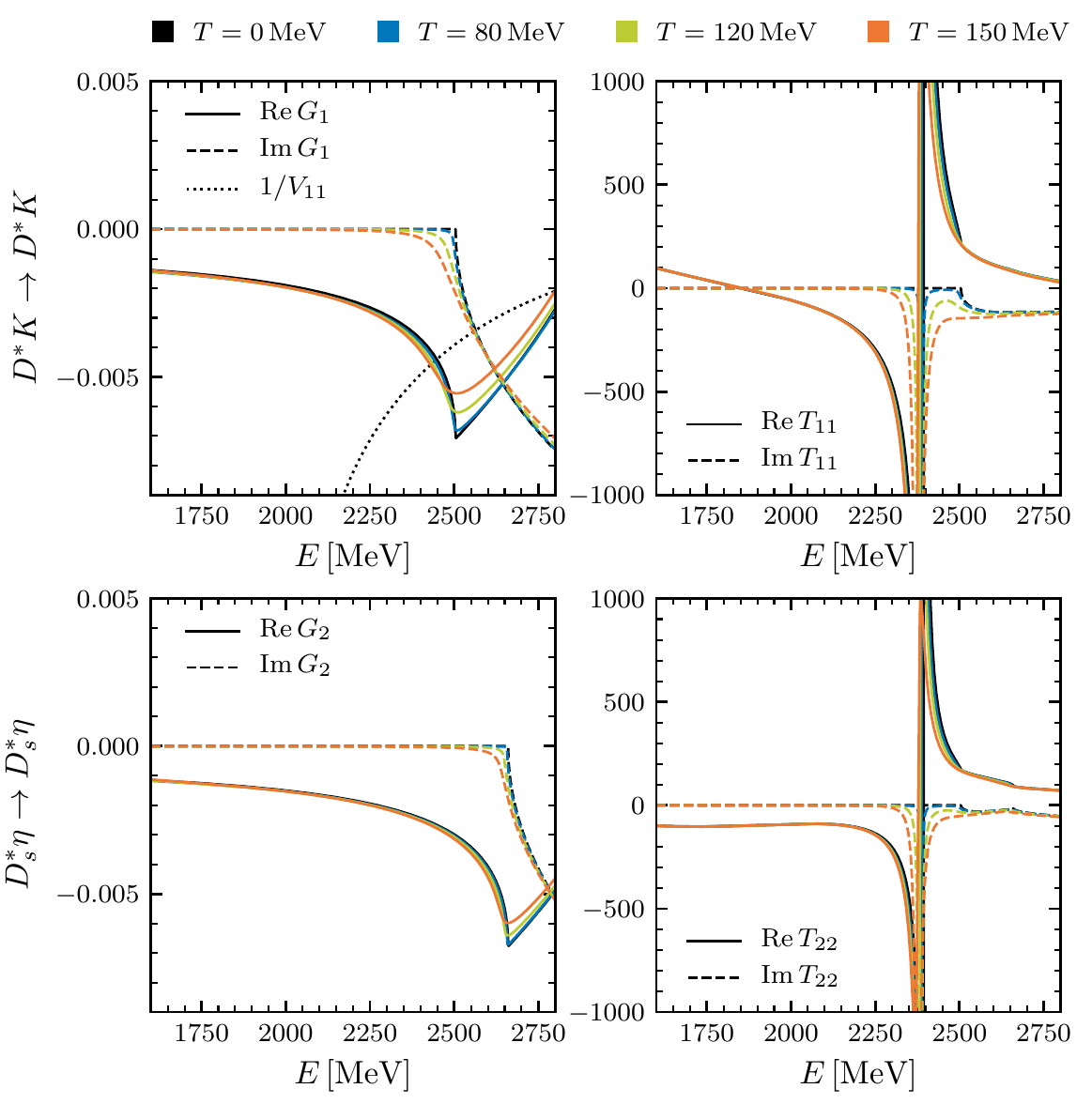}
   \caption{The same as in Fig.~\ref{fig:VGT_0_05_vec} in the sector with $(S,I)=(1,0)$. The subindices 1, 2 refer to the channels $D_s^*\pi$ and $D^*K$, respectively.}
   \label{fig:VGT_1_0_vec}
   \end{figure}

\subsubsection{$J=1$ case: Interactions and $D_1^*(2430)$ and $D_{s1}^*(2460)$}

The coupled-channels interaction of the pseudoscalar meson octet with the heavy vector mesons gives rise to a very similar phenomenology to that found for the interaction with the heavy pseudoscalars, only displaced toward higher energies by the increase of the thresholds due to the mass difference between the vector and pseudoscalar heavy mesons. This is clearly seen in Figs.~\ref{fig:VGT_0_05_vec} and \ref{fig:VGT_1_0_vec}, where the loop functions and the diagonal amplitudes of the $J^P=1^+$ interaction in the $(S,I)=(0,\frac12)$ and $(1,0)$ sectors are shown, respectively, as functions of the total energy for a total momentum $\vec{P}=0$ and various temperatures.

At $T=0$ we also find five poles in the $J=1$ amplitudes, which are  summarized in Table~\ref{tab:poles1}. As seen, a double pole structure, which can be identified with the $D_1^*(2430)$ resonance listed in the PDG \cite{Tanabashi:2018oca}, is obtained in the $(S,I)=(0,\frac12)$ sector, one pole being mostly a ${D^*\pi}$ state and the other mostly a ${D_s^*\bar{K}}$ one. A resonance coupling mostly to ${D^*K}$ states is obtained in the $(S,I)=(1,0)$ sector, although the present model locates it at a somewhat lower energy than the mass of the $D_{s1}^*(2460)$ of the PDG\cite{Tanabashi:2018oca}. A virtual state is found in the $(S,I)=(-1,0)$ sector, while a resonance with   $(S,I)=(1,1)$, having no PDG counterpart, is also seen.

We note that the cutoff dependence analyses performed in Refs.\cite{Montana:2017kjw,Ramos:2020bgs} indicate that employing higher (lower) values of the cutoff lowers (increases) the energies of the dynamically generated states, due to the larger (smaller) amount of phase space included in the unitarized amplitudes.  In our case, when the cutoff value is varied between 600 and 1000 MeV, the mass of the resonances in the $(S, I) = (0, 1/2)$ sector, $D^*_0(2300)$ and $D^*_1(2430)$, both change moderately by 
$^{+5}_{-15}$ MeV, while a larger change, of $\pm 70$ MeV, is observed for both bound states in the $(S, I) = (1, 0)$ sector, $D^*_{s0}(2317)$ and $D^*_{s1}(2460)$.

\begin{table}[htbp!]
\begin{tabular}{ccccccc}
\hline
& $(S,I)$ & RS & $M_R$ & $\Gamma_R/2$ & $|g_i|$ & $\chi_i$ \\
& & & (MeV) & (MeV) & (GeV) &  \\
\hline
\hline
 & $(-1,0)$ & $(-)$ & $2404.9$ & $87.8$ & $|g_{D^*\bar{K}}|=13.2$ & $\chi_{D^*\bar{K}}=0.53$ \\
\hline
$D_1^*(2430)$ & $(0,\frac12)$ & $(-,+,+)$ & $2222.3$ & $84.7$  & $|g_{D^*\pi}|=9.5$  & $\chi_{D^*\pi}=0.40$ \\
 &    &           &        &       & $|g_{D^*\eta}|=0.4$  & $\chi_{D^*\eta}=0.00$ \\
 &   &           &        &       & $|g_{D_s^*\bar{K}}|=5.7$  & $\chi_{D_s^*\bar{K}}=0.05$ \\
 &  & $(-,-,+)$ & $2654.6$ & $117.3$ & $|g_{D^*\pi}|=6.5$  & $\chi_{D^*\pi}=0.09$ \\
 &   &           &        &       & $|g_{D^*\eta}|=10.0$  & $\chi_{D^*\eta}=0.40$ \\
 &   &           &        &       & $|g_{D_s^*\bar{K}}|=18.5$ & $\chi_{D_s^*\bar{K}}=1.47$ \\
    \hline
$D_{s1}^*(2460)$ &   $(1,0)$ & $(+,+)$ & $2393.3$ & $0.0$ & $|g_{D^*K}|=14.2$ & $\chi_{D^*K}=0.68$ \\
 &    &           &        &       & $|g_{D_s^*\eta}|=9.7$ & $\chi_{D_s^*\eta}=0.17$ \\
\hline
 & $(1,1)$ & $(-,-)$ & $2392.2$ & $193.0$ & $|g_{D_s^*\pi}|=7.9$ & $\chi_{D_s^*\pi}=0.22$ \\
 &         &         &          &         & $|g_{D^*K}|=6.3$     & $\chi_{D^*K}=0.08$ \\
\hline
 \end{tabular}
 \centering
 \caption{Dynamically generated poles in the $J=1$ sector. Table with a similar structure than the one in Table~\ref{tab:poles0}.}
 \label{tab:poles1}
 \end{table}

\subsection{Open-charm spectral functions and dynamically generated states at $T \neq 0$}

We conclude the presentation of our results by analyzing in this section the finite-temperature effects on the interaction of the open-charm mesons with the light pseudoscalars, together with the consequences for the ground-state spectral functions and the dynamically generated states at finite temperature. In particular, we describe the thermal dependences of the masses and widths for open-charm ground and excited states for temperatures below the chiral restoration temperature $T_\chi=156$ MeV~\cite{Aoki:2006we}.

\subsubsection{Interactions and open-charm spectral functions}

The inclusion of temperature on the real and imaginary parts of the different meson-meson loop functions results in a smoothening of the real and imaginary parts of the scattering matrices in all strange-isospin sectors for both pseudoscalar and vectors mesons, as seen in Figs.~\ref{fig:VGT_0_05}--\ref{fig:VGT_1_0_vec}. As the meson Bose-Einstein distributions in the intermediate meson-meson propagators extend over higher momenta for larger temperatures, sharp meson-meson thresholds are diluted and the strength of the real and imaginary parts of the loop functions is smoothened out. As a consequence, the corresponding scattering amplitudes are smeared out while spreading over a wider energy range. Physically, this can be understood as larger temperatures result in a larger available phase space for decay.

The smearing of the scattering amplitudes with temperature  gives rise to the broadening of the spectral functions, as seen in Figs.~\ref{fig:SpectralFunction_ps} and \ref{fig:SpectralFunction_vec}. In Fig.~\ref{fig:SpectralFunction_ps} we show the spectral functions in a pionic bath as a function of the meson energy for different temperatures up to $T=150$ MeV in the case of the pseudoscalar open-charm ground-state mesons, that is, $D$ (left panel) and $D_s$ (right panel), whereas in Fig.~\ref{fig:SpectralFunction_vec} we display the  vector open-charm ground-state spectral functions for $D^*$ (left panel) and $D_s^*$(right panel) in a pionic medium, respectively. We clearly see the increased broadening of all spectral functions with temperature. Moreover, we observe that the maximum of the spectral functions is shifted toward lower energies for higher temperatures, indicating the attractive character of the interaction of open-charm mesons with a pionic bath. 

It is also interesting to note the similar shape of the pseudoscalar and vector open-charm ground-state spectral functions, that is, the parallel behavior with temperature of the $D$ and $D^*$ spectral functions as well as the $D_s$ and $D^*_s$ ones. As previously mentioned, at LO in the heavy-mass expansion, pseudoscalar and vector open-charm ground states are related by heavy-quark spin symmetry. Therefore, a similar behavior with temperature of the spectral functions in the pseudoscalar and vector sectors is expected.

  \begin{figure}[htbp!]
   \includegraphics[width=0.8\textwidth]{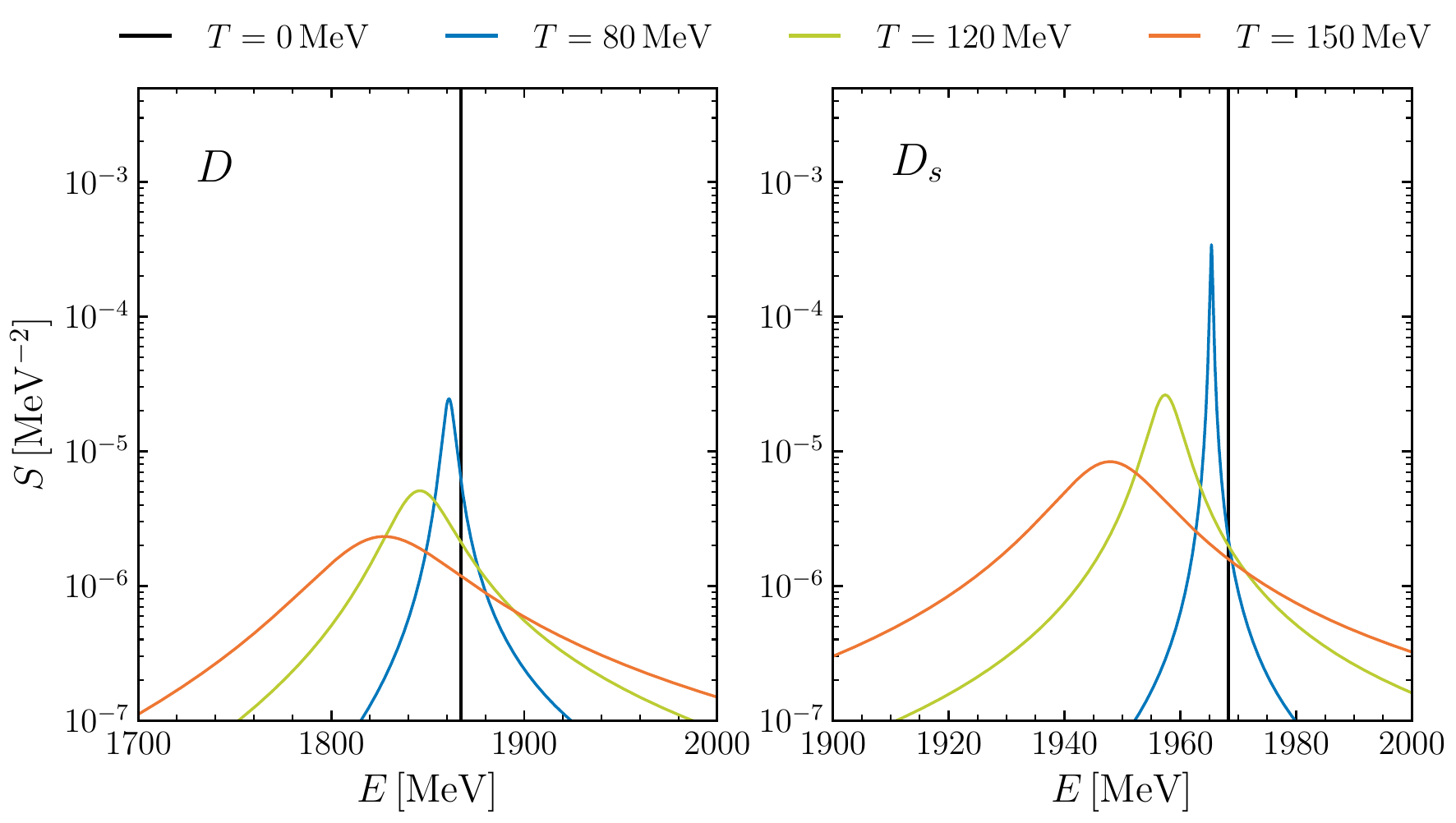}
   \caption{Spectral functions of the $J=0$ ground states ($D$ and $D_s$) at different temperatures in a pionic bath.}
   \label{fig:SpectralFunction_ps}
   \end{figure}
   
    \begin{figure}[htbp!]
   \includegraphics[width=0.8\textwidth]{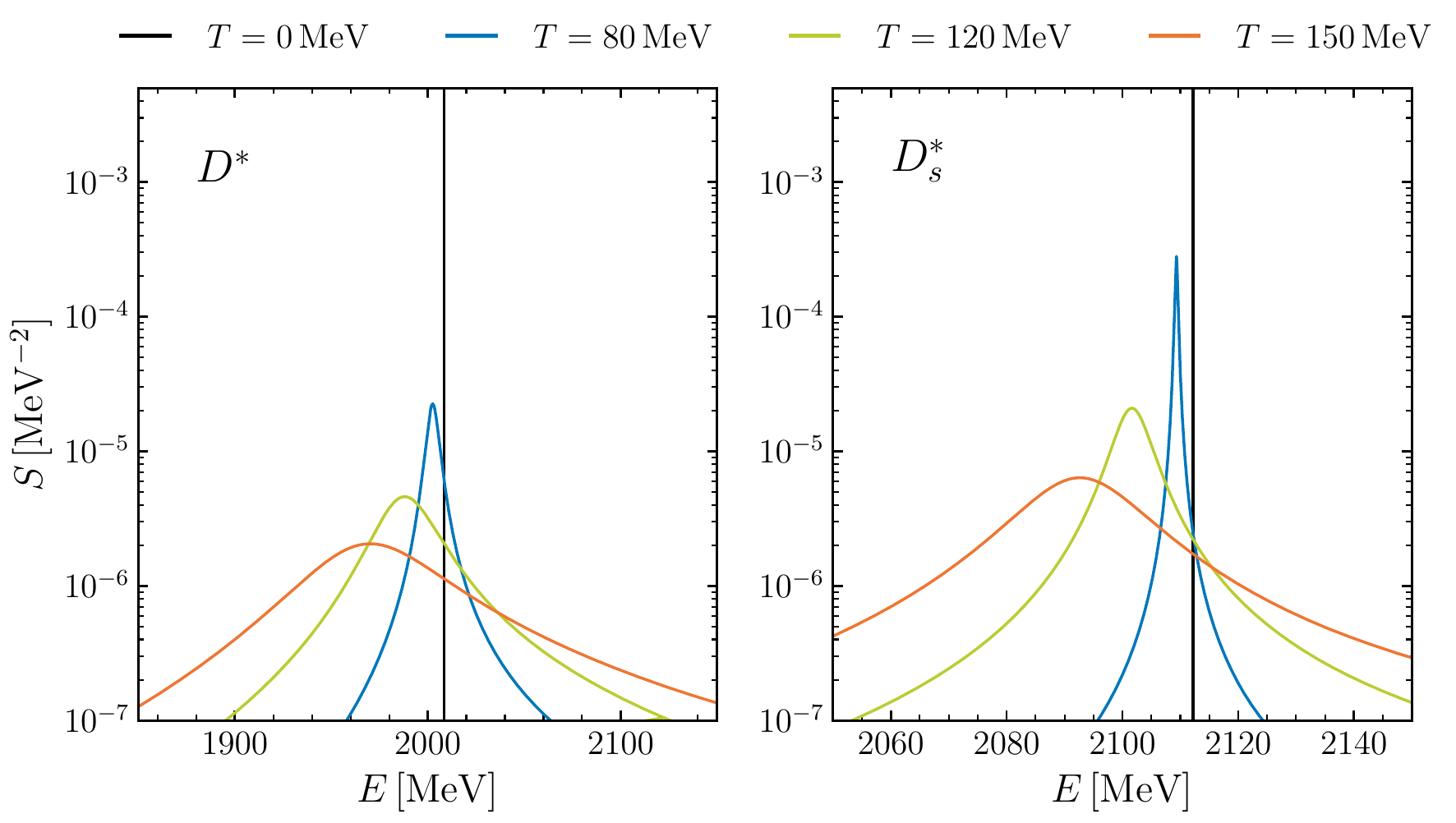}
   \caption{Spectral functions of the $J=1$ ground states ($D^*$ and $D_s^*$) at different temperatures in a pionic bath.}
   \label{fig:SpectralFunction_vec}
   \end{figure}

\subsubsection{Determination of masses and widths at $T\neq 0$~\label{sec:thermalmasses}}

In order to obtain a quantitative description of the thermal dependence of the masses and widths of the open-charm ground states, one should analyze  the behavior of the corresponding spectral functions with temperature. The mass change with temperature can be extracted from the position of the peak of the spectral function at different temperatures, that is, from the position of the so-called quasiparticle peak, $\omega_{qp}$, obtained from: 
\begin{equation}
\omega_{qp}^2-\vec{q}\,^2-m^2_D-{\rm Re} \Pi(\omega_{qp},\vec{q};T)=0 ,
\end{equation}
whereas the variation of the width with temperature can be obtained from the thermal spectral function at half height. 

However, the determination of the behavior of the mass and width with temperature of the dynamically generated states, such as $D^*(2300)$ and $D_{s0}^* (2317)$ as well as $D_1^*(2430)$ and $D_{s1}^*(2460)$, is rather delicate. The calculation of the poles in the complex plane at finite temperature, while performing the self-consistency program, is computationally very expensive and unfeasible. In addition, the analytic continuation to the different RSs should be performed, not from the real energy axis, but from the imaginary Matsubara frequencies, and it is not clear how to perform this technically. Therefore, we employ the method described in the following to obtain the particle properties on the real axis, through fits of the imaginary part of the unitarized scattering amplitudes at finite temperature shown in Figs.~\ref{fig:VGT_0_05}-\ref{fig:VGT_1_0_vec}.

For isolated resonances close to the real energy axis and not close to any threshold we simply use a Breit-Wigner form.  But in the case of resonances that interact with the background of another resonance (in the coupled-channel case) we use a Breit-Wigner-Fano shape~\cite{Fano:1961zz}. This can be used for the lower pole in the double pole structure of the $D_0^*(2300)$. Indeed, we have checked that the obtained mass and width of the fit at $T=0$ are in very good agreement with the values of the pole mass and the width in Table~\ref{tab:poles0}.

The Breit-Wigner-Fano-type distribution provides a simple parametrization to describe the distorted lower resonance at finite temperature:
\begin{equation}\label{BreitWignerFano}
 f^{\rm BWF}(E;A,m_R,\Gamma_R,q)=A\frac{\Gamma_R/2+ (E-m_R)/q}{(\Gamma_R/2)^2+(E-m_R)^2} \ ,
\end{equation}
where $q$ is the Fano parameter measuring the ratio of resonant scattering to background scattering amplitude. In the absence of background the value of $q$ goes to infinite and Eq.~(\ref{BreitWignerFano}) becomes the usual Breit-Wigner distribution.

For resonances close to a threshold we fit to a Flatt\'e-type distribution~\cite{Flatte:1976xu}. In particular for the higher pole in the double pole structure we first subtract the background and then, we use a generalized Flatt\'e parametrization with three coupled-channels:
\begin{eqnarray}\label{Flatte3channels} \nonumber
 {\rm Im\,} T_{ij}(s;C,m_R,g_1,g_2,g_3) &=& Cg_ig_j\Bigg[\frac{\rho_1g_1^2}{\big(m_R^2-s+|\rho_2|g_2^2+|\rho_3|g_3^2\big)^2+\big(\rho_1g_1^2\big)^2} \ \theta(m_D+m_\eta-\sqrt{s}) \\ \nonumber
 &+& \frac{\rho_1g_1^2+\rho_2g_2^2}{\big(m_R^2-s+|\rho_3|g_3^2\big)^2+\big(\rho_1g_1^2+\rho_2g_2^2\big)^2} \ \theta(\sqrt{s}-m_D-m_\eta) \ \theta(m_{D_s}+m_{\bar{K}}-\sqrt{s}) \\
 &+& \frac{\rho_1g_1^2+\rho_2g_2^2+\rho_3g_3^2}{\big(m_R^2-s\big)^2+\big(\rho_1g_1^2+\rho_2g_2^2+\rho_3g_3^2\big)^2} \ \theta(\sqrt{s}-m_{D_s}-m_{\bar{K}})\Bigg] \ ,
\end{eqnarray}
where $\rho_i$ stands for the phase space of the $i{\rm th}$ channel,
\begin{equation}
 \rho_i(\sqrt{s})=\frac{2p_i(\sqrt{s})}{\sqrt{s}}=\Bigg[\Bigg(1-\frac{(m_{ia}+m_{ib})^2}{s}\Bigg)\Bigg(1-\frac{(m_{ia}-m_{ib})^2}{s}\Bigg)\Bigg]^{1/2} \ .
\end{equation}
The resonance width is given by
\begin{equation}\label{widthFlatte}
 m_R\Gamma_R=\rho_1(m_R)g_1^2+\rho_2(m_R)g_2^2+\rho_3(m_R)g_3^2 \ ,
\end{equation}
where the phase spaces have been evaluated at the resonance mass.
In our case, the subindices correspond to $1\equiv D\pi$, $2\equiv D\eta$, and $3\equiv D_s\bar{K}$. In order to avoid an ill behavior of the fit due to the large amount of free parameters, the value of $g_1$ is imposed to vary linearly from its lowest value at $T=0$ to the highest one at $T=150$ MeV.

The Breit-Wigner-Fano distribution is also used for isolated resonances at high temperatures, if they become wide enough to be affected by threshold effects. 

\subsubsection{$J=0$ case: $D$/$D^*(2300)$ and  $D_s$/$D_{s0}^*(2317)$ }

Our results for the masses and widths of the ground-state mesons $D$ and $D_s$ as well as the dynamically generated states $D^*(2300)$ and $D_{s0}^*(2317)$ are shown in Fig.~\ref{fig:resonances_ps} for a pionic bath (solid lines) and when the medium is populated by  both pions and kaons (dashed lines). They are summarized as follows:

\begin{enumerate}
 \item In a pionic medium the ground-state $D$ mass shows a sizable decrease of $\Delta m_D \sim 40$ MeV at the largest temperature $T=150$ MeV. This reduction is twice larger than that observed in~\cite{Fuchs:2004fh}, where a more phenomenological approach for the $D$-meson propagator is used. Our reduction, however, is smaller than the one shown in Ref.~\cite{Sasaki:2014asa}, where a nonunitarized ChPT is considered. In contrast, the $SU(4)$ effective approach of \cite{Cleven:2017fun} shows no significant modification of the mass of the ground state.  On the other hand, the mass of the two poles of the $D^*_0(2300)$ change less rapidly with temperature compared to the ground state, moving downward and distancing from each other with increasing temperature. As a consequence, we cannot conclude that masses of opposite parity states become degenerate for the temperatures studied, as discussed in Ref.~\cite{Montana:2020lfi}.
  
 \item The width of the nonstrange states increases with temperature, being more relevant the change in width for the ground state. The $D$ meson has a width of around $\sim70$ MeV at $T =150$ MeV, consistent with~\cite{Cleven:2017fun,Fuchs:2004fh,He:2011yi}. The widths of the two poles of the $D_0^*(2300)$ increase moderately with respect to the vacuum values. 

 \item In the strangeness sector the $D_s$ and $D_{s0}^*(2317)$ decrease their masses with temperature in a similar manner. Thus, both chiral partners are still far from chiral degeneracy at $T=150$ MeV. The trend of the mass of both states is in line with the low temperature behavior observed in Ref.~\cite{Sasaki:2014asa} within the linear sigma model.

 \item The decay widths of both strange partners increase similarly with the temperature of the pionic bath, reaching moderate values of $15-20$ MeV at $T=150$ MeV.  Compared to the ground state, the $D_{s0}^* (2317)$ contains the additional contribution of the decay into $DK$ states due to the reduction of the mass and the widening of the $D$ meson.
 
 \item In a bath that includes $K$ and $\bar{K}$ in addition to pions (see dashed lines in Fig.~\ref{fig:resonances_ps}), the masses of the ground states $D$ and $D_s$ decrease an additional amount of 5 MeV and 4 MeV at $T=150$ MeV, respectively. The modification of the width is, however, different for the nonstrange and the strange states. In the case of the $D$ meson, the width increases around 20\% while the one for the $D_s$ meson is almost 3 times larger than the width in a pionic medium at $T=150$ MeV. This follows from the stronger interaction of the $D_s$ mesons with kaons than with pions. On the other hand, the effect of the pionic and kaonic bath on the dynamically generated states is rather moderate, with the masses of the two resonances of the $D_0^*(2300)$ increasing a few MeV and no significant modification of the widths, whereas the mass of the $D_{s0}^*(2317)$ drops slightly and the width increases around 5 MeV.
 
\end{enumerate}

\begin{figure}[htp!]
\includegraphics[width=\textwidth]{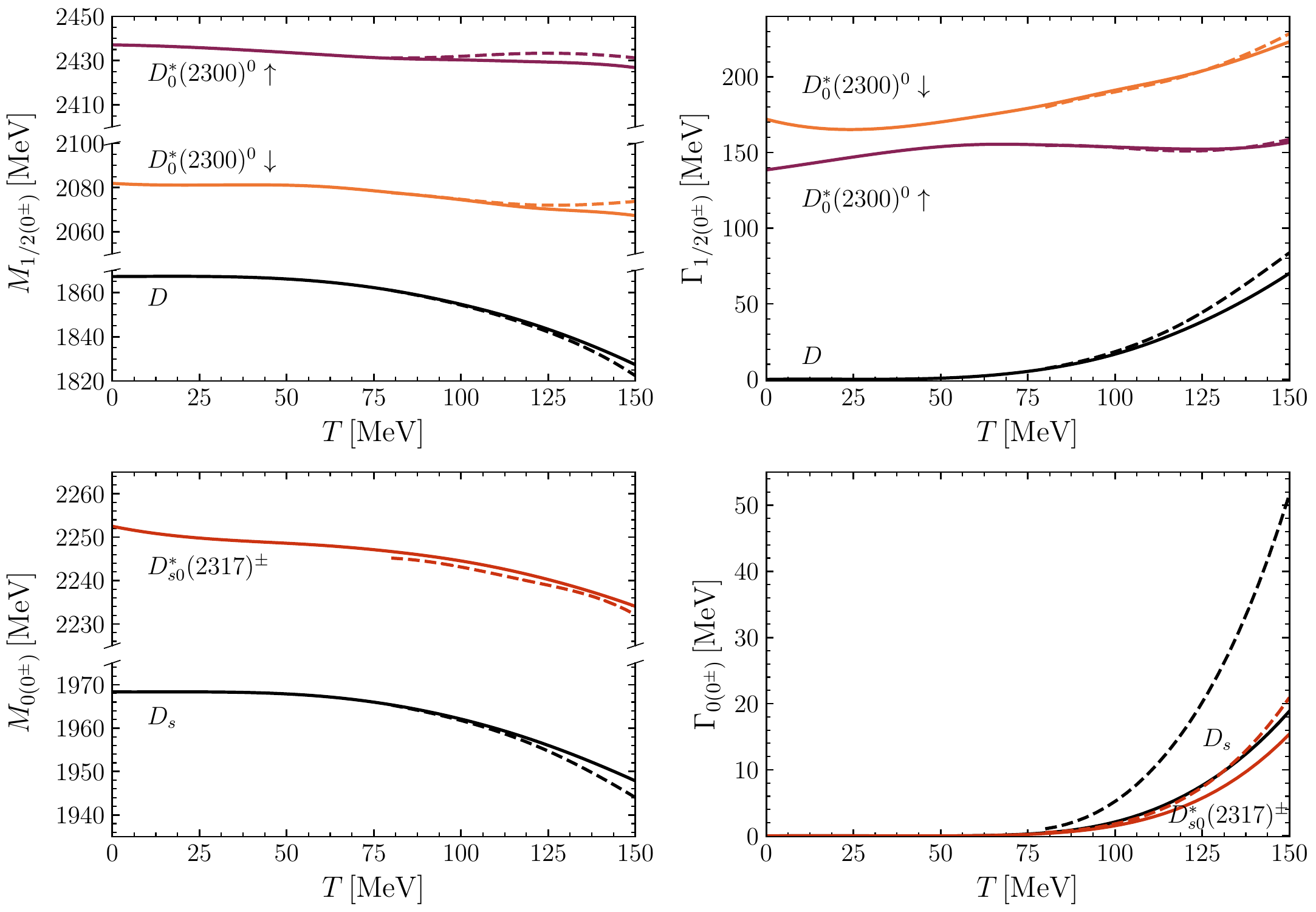}
\caption{Temperature evolution of the mass (left panels) and width (right panels) of the $J=0$ ground-state mesons and dynamically generated states in the $(S;I) = (0; \frac{1}{2})$ sector (upper panels) and in the $(S;I) = (1;0)$ sector (lower panels) in a pionic medium (solid lines) and in a medium with $\pi$, $K$ and $\bar{K}$ (dashed lines)}.
\label{fig:resonances_ps}
\end{figure}

\subsubsection{$J=1$ case: $D^*$/$D_1^*(2430)$ and  $D^*_s$/$D_{s1}^*(2460)$}

With regards to the thermal evolution of the masses and widths in the $J=1$ case for the ground-state mesons $D^*$ and $D^*_s$ as well as the dynamically generated states $D_1^*(2430)$ and $D_{s1}^*(2460)$, we observe a clear parallelism in their behavior with that obtained for the $J=0$ states, and which is shown in Fig.~\ref{fig:resonances_vec}. Again, this is due to the fact that interactions of light mesons with pseudoscalar open-charm ones are related by heavy-quark spin symmetry to those with vector open-charm ground states. Thus, our conclusions are similar to the ones presented for $J=0$, namely, the following:

\begin{enumerate}

 \item In a medium with pions, the  $D^*$ mass shows a sizable decrease of $\Delta m_{D^*} \sim 40$ MeV at the largest temperature $T=150$ MeV, similar to the $D$ mass shift.  As for  two poles that formed the $D_1^*(2430)$, their masses decrease less rapidly with temperature compared to the ground state, distancing from each other as temperature increases, in an analogous manner as for the two poles of the  $D^*_0(2300)$. As a consequence, also in the $J=1$ case, we cannot conclude that masses of opposite parity states become degenerate with temperature, at least for the range of temperatures studied here.  
 
 \item The width of the nonstrange states increases with temperature, being the change more relevant for the ground state, as in the case of $J=0$. The $D^*$ meson shows a similar width of around $\sim70$ MeV at $T =150$ MeV as the $D$. The widths of the two poles of the $D_1^*(2430)$ increase moderately with respect to vacuum value, as seen for the $D^*_0(2300)$.

 \item In the strangeness sector, the $D^*_s$ and its chiral partner $D_{s1}^*(2460)$ behave similarly to the $D_s$ and $D_{s0}^*(2317)$ states, as they decrease their masses with temperature analogously. Hence, the $D^*_s$ and $D_{s1}^*(2460)$  states follow the trend of the $D_s$ and $D_{s0}^*(2317)$, so they are still far from chiral degeneracy at $T=150$ MeV. 
 
 \item The decay widths of both strange partners increase with temperature and reach similar values at $T=150$ MeV, becoming slightly larger compared to the  ones of  the $D_s$ and $D_{s0}^*(2317)$.  Similarly to the $D_{s0}^*(2317)$, the width of $D_{s1}^*(2460)$ is affected by the reduction of the mass and the widening of the $D^*$ due to the dominant contribution of the $D^* K$ channel in its dynamical generation.
 
 \item The addition of kaons in the mesonic bath (see dashed lines in Fig.~\ref{fig:resonances_vec}) results in a modification of the ground-state $D^*$ and $D_s^*$ masses and widths in a similar manner to that of the $D$ and $D_s$. Their masses drop 5 and 4 MeV and their widths increase about 20\% and by almost a factor 3 at $T=150$ MeV, respectively. Furthermore the modification of the dynamically generated states $D_1^*(2430)$ and $D_{s1}^*(2460)$ is analogous to that described above for the case of the $J=0$ states.

\end{enumerate}

\begin{figure}[htp!]
\includegraphics[width=\textwidth]{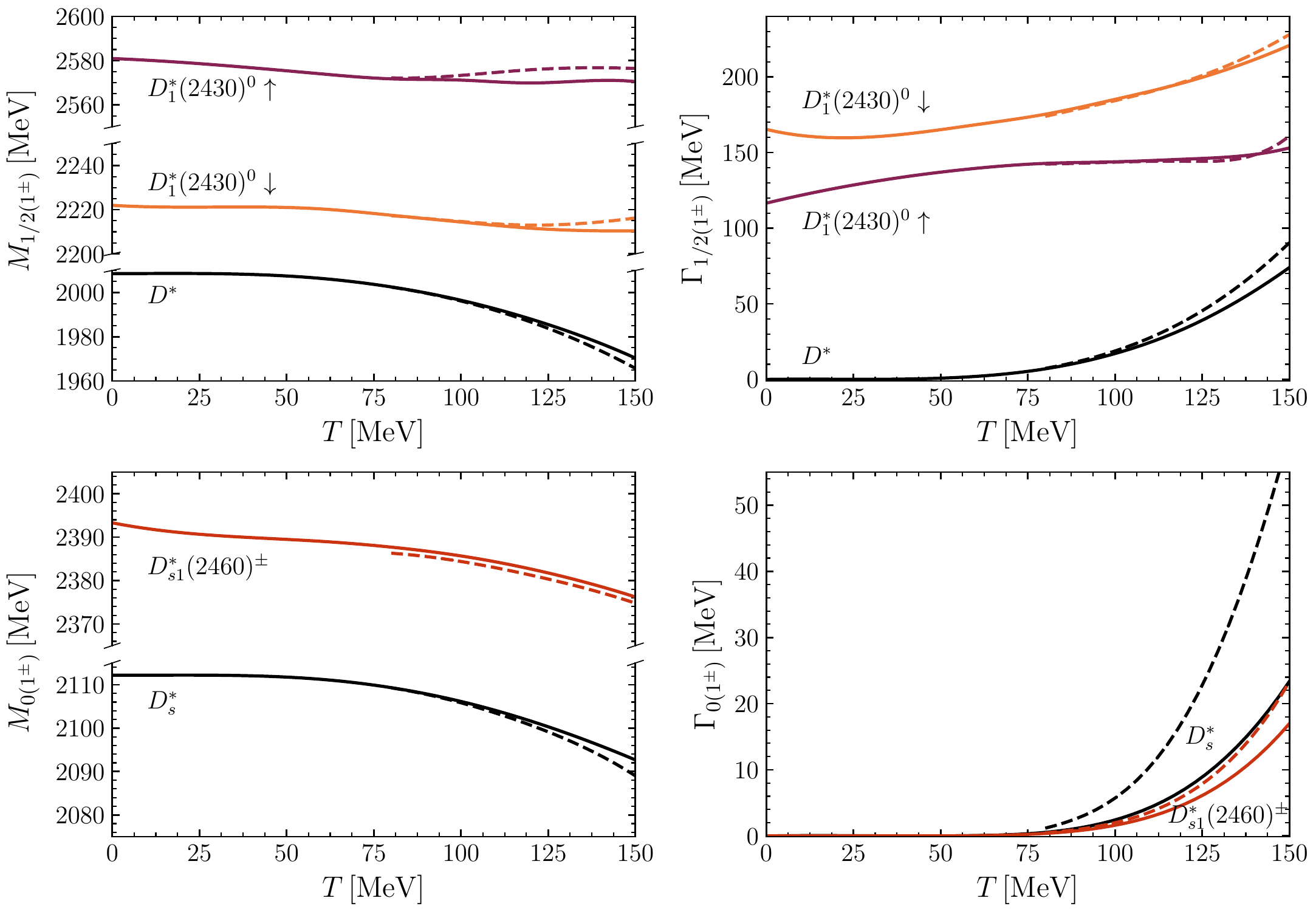}
\caption{Temperature evolution of the mass (left panels) and width (right panels) of the $J=1$ ground-state mesons and dynamically generated states in the $(S, I) = (0, \frac{1}{2})$ sector (upper panels) and in the $(S, I) = (1, 0)$ sector (lower panels) in a pionic medium (solid lines) and in a medium with $\pi$, $K$ and $\bar{K}$ (dashed lines)}.
\label{fig:resonances_vec}
\end{figure}


\newpage
\section{Conclusions and Outlook}
\label{conclusions}

In this work we have studied the properties of pseudoscalar and vector charm mesons in a thermal medium up to $T=150$ MeV. The interactions of the ground-state $D, D_s$ and $D^*, D_s^*$ with light mesons ($\pi, K, \bar{K}, \eta$) are described by an effective field theory based on chiral and heavy-quark spin symmetries. The medium modification of the heavy-meson propagator is calculated in a self-consistent way, in which the charm meson self-energy is corrected by the effects of the $T$-matrix amplitude, which in turn, is computed via the solution of a Bethe-Salpeter equation, with in-medium heavy-meson propagators.

In vacuum, we have analyzed the different dynamical generated states of the $T$-matrix in a series of channels by extending the scattering amplitudes to the appropriate Riemann sheet. In the $J=0$ case we have described the $D^*(2300)$ state and its double-pole structure, the $D_{s0}^* (2317)$ bound state, together with a virtual state and an additional resonance not yet experimentally identified. In $J=1$, we have paid attention to the $D_1^*(2430)$ resonance (also with a double pole structure), the $D_{s1}^*(2460)$, as well as  a virtual state and a resonance. The parallelism between the $J=0$ and $J=1$ sectors is due to the heavy-quark spin symmetry, which is maintained at LO, and only broken due to the explicit value of the heavy-meson vacuum masses.

At finite temperature we have studied the thermal dependence of the ground-state $D$, $D^*$, $D_s$ and $D_s^*$ mesons, as well as the experimentally identified dynamically generated states. We have observed a generic downshift of the thermal masses with temperature, as large as $\Delta m_D,\Delta m_{D^*} \simeq 40$ MeV for the nonstrange mesons at $T=150$ MeV in a pionic bath. Their strange counterparts, $D_s$ and $D_s^*$ present half of this mass downshift at the same temperature. We have also obtained that the decay widths of these states increase with temperature up to values of $\Gamma_D,\Gamma_{D^*} \simeq 70$ MeV, and $\Gamma_{D_s},\Gamma_{D_s^*} \simeq 20$ MeV at $T=150$ MeV if we consider a pionic medium. The mass of both poles of the $D_0^*(2300)$ resonance as well as that of the $D_{s0}^* (2317)$ decreases more softly with temperature with respect to the ground states in this pionic environment. On the other hand, the decay widths in pionic matter increase in a similar magnitude as those of $D$ and $D_s$, with the exception of the lower pole of the $D_0^*(2300)$ resonance, with an increase of $\Delta \Gamma=50$ MeV with respect to its vacuum value.  The addition of kaons in the meson bath results in a slight modification of the masses of the nonstrange and strange $J=0$ ground-state mesons, whereas the widths increase significantly, in particular for the $D_s$. The effect of the kaonic and pionic medium on the masses and widths of the dynamically generated states is, however, rather moderate. We notice that the modifications of the $J=1$ states are similar to the $J=0$ sector due to heavy-quark spin symmetry.

Our results provide the first systematic approach to the thermal effects on open-charm mesons below the crossover temperature in view of the present and future heavy-ion experiments at low-baryonic densities as well as forthcoming results in lattice-QCD simulations at finite temperature. In the future we will explore the effects of medium-modified light mesons in the self-consistent calculation, and the straightforward  extension to bottom flavor, exploiting the heavy-quark flavor symmetry together with the heavy-quark spin symmetry. On the other hand, we will apply our results of the $D$-meson spectral functions to the calculation of transport coefficients like the heavy-flavor diffusion coefficient. This will constitute an extension of our previous results~\cite{Tolos:2013kva,Ozvenchuk:2014rpa,Song:2015sfa,Tolos:2016slr,Das:2016llg} incorporating off-shell effects in the kinetic approach.

\begin{acknowledgments}

J. M. T.-R. acknowledges the hospitality of the Institut de Ci\`encies de l'Espai (CSIC) and the Universitat de Barcelona, where part of this work was carried out. He thanks discussion with \'A. G\'omez-Nicola and J.A. Oller on the subject.

G.M. and A.R.  acknowledge  support from the Spanish Ministerio de Econom\'ia y Competitividad (MINECO) under the project MDM-2014-0369 of ICCUB (Unidad de Excelencia ``Mar\'ia de Maeztu''), and, with additional European FEDER funds, under the Contract No. FIS2017-87534-P. G.M. also acknowledges support from the FPU17/04910 Doctoral Grant from the Spanish Ministerio de Educaci\'on, Cultura y Deporte. L.T. acknowledges support from the FPA2016-81114-P Grant from the former Ministerio de Ciencia, Innovaci\'on  y  Universidades,  the PID2019-110165GB-I00 Grant from the Ministerio de Ciencia e Innovaci\'on, the Heisenberg  Programme  of the Deutsche Forschungsgemeinschaft (DFG, German research Foundation) under the Project No. 383452331 and the THOR COST Action CA15213. L.T. and J. M. T.-R. acknowledge support from the DFG through Projects no. 411563442 (Hot Heavy Mesons) and No. 315477589 - TRR 211 (Strong-interaction matter under extreme conditions). We also thank the EU STRONG-2020 project under the program  H2020-INFRAIA-2018-1, Grant No. 824093.

\end{acknowledgments}

\appendix

\section{Isospin coefficients in the charge basis}
\label{app:isospincoeff}

In this appendix we provide the isospin coefficients in the charge basis. They are shown in Tables~\ref{tab:coeff_Qbasis} and~\ref{tab:coeff_Qbasis2}.
  
\newpage 
\begin{table}[H]
\begin{tabular}{c l | c c c c c}
\hline
 $(S,Q)$  &  Channel & $C_{\rm LO}^{jk}$ & $C_0^{jk}$ & $C_1^{jk}$ &  $C_{\rm 24}^{jk}$ & $C_{35}^{jk}$  \\  
\hline
 $(-1,-1)$ & $D^0K^-\rightarrow D^0K^-$ & $1$ & $m_K^2$ & $-m_K^2$ & $1$ & $1$ \\
\hline
 $(-1,0)$  & $D^0\bar{K}^0\rightarrow D^0\bar{K}^0$ & $0$ & $m_K^2$ & $0$ & $1$ & $0$ \\
   & $D^0\bar{K}^0\rightarrow D^+K^-$ & $1$ & $0$ & $-m_K^2$ & $0$ & $1$ \\
   & $D^+K^-\rightarrow D^+K^-$ & $0$ & $m_K^2$ & $0$ & $1$ & $0$ \\
\hline
 $(-1,+1)$  & $D^+\bar{K}^0\rightarrow D^+\bar{K}^0$ & $1$ & $m_K^2$ & $-m_K^2$ & $1$ & $1$ \\
\hline
 $(0,-1)$ & $D^0\pi^-\rightarrow D^0\pi^-$ & $1$ & $m_\pi^2$ & $-m_\pi^2$ & $1$ & $1$ \\
\hline
 $(0,0)$   & $D^0\pi^0\rightarrow D^0\pi^0$ & $0$ & $m_\pi^2$ & $-m_\pi^2$ & $1$ & $1$ \\
   & $D^0\pi^0\rightarrow D^+\pi^-$ & $-\sqrt{2}$ & $0$ & $0$ & $0$ & $0$ \\
   & $D^0\pi^0\rightarrow D_s^+K^-$ & $-\frac{1}{\sqrt{2}}$ & $0$ & $-\frac{1}{2\sqrt{2}}(m_K^2+m_\pi^2)$ & $0$ & $\frac{1}{\sqrt{2}}$ \\
   & $D^0\pi^0\rightarrow D^0\eta$ & $0$ & $0$ & $-\frac{1}{\sqrt{3}}m_\pi^2$ & $0$ & $\frac{1}{\sqrt{3}}$ \\ 
   & $D^+\pi^-\rightarrow D^+\pi^-$ & $-1$ & $m_\pi^2$ & $-m_\pi^2$ & $1$ & $1$ \\
   & $D^+\pi^-\rightarrow D_s^+K^-$ & $-1$ & $0$ & $-\frac{1}{2}(m_K^2+m_\pi^2)$ & $0$ & $1$ \\ 
   & $D^+\pi^-\rightarrow D^0\eta$ & $0$ & $0$ & $-\sqrt{\frac{2}{3}}m_\pi^2$ & $0$ & $\sqrt{\frac{2}{3}}$ \\    
   & $D_s^+K^-\rightarrow D_s^+K^-$ & $-1$ & $m_K^2$ & $-m_K^2$ & $1$ & $1$ \\ 
   & $D_s^+K^-\rightarrow D^0\eta$ & $-\sqrt{\frac{3}{2}}$ & $0$ & $\frac{1}{2\sqrt{6}}(5m_K^2-3m_\pi^2)$ & $0$ & $-\frac{1}{\sqrt{6}}$ \\
   & $D^0\eta\rightarrow D^0\eta$ & $0$ & $m_\eta^2$ & $-\frac{1}{3}m_\pi^2$ & $1$ & $\frac{1}{3}$ \\ 
\hline
 $(0,+1)$   & $D^0\pi^+\rightarrow D^0\pi^+$ & $-1$ & $m_\pi^2$& $-m_\pi^2$ & $1$ & $1$ \\
   & $D^0\pi^+\rightarrow D^+\pi^0$ & $\sqrt{2}$ & $0$ & $0$ & $0$ & $0$ \\
   & $D^0\pi^+\rightarrow D_s^+\bar{K}^0$ & $-1$ & $0$ & $-\frac{1}{2}(m_K^2+m_\pi^2)$ & $0$ & $1$ \\
   & $D^0\pi^+\rightarrow D^+\eta$ & $0$ & $0$ & $-\sqrt{\frac{2}{3}}m_\pi^2$ & $0$ & $\sqrt{\frac{2}{3}}$ \\
   & $D^+\pi^0\rightarrow D^+\pi^0$ & $0$ & $m_\pi^2$ & $-m_\pi^2$ & $1$ & $1$ \\
   & $D^+\pi^0\rightarrow D_s^+\bar{K}^0$ & $\frac{1}{\sqrt{2}}$ & $0$ & $\frac{1}{2\sqrt{2}}(m_K^2+m_\pi^2)$ & $0$ & $-\frac{1}{\sqrt{2}}$ \\
   & $D^+\pi^0\rightarrow D^+\eta$ & $0$ & $0$ & $\frac{1}{\sqrt{3}}m_\pi^2$ & $0$ & $-\frac{1}{\sqrt{3}}$ \\
   & $D_s^+\bar{K}^0\rightarrow D_s^+\bar{K}^0$ & $-1$ & $m_K^2$ & $-m_K^2$ & $1$ & $1$ \\
   & $D_s^+\bar{K}^0\rightarrow D^+\eta$ & $-\sqrt{\frac{3}{2}}$ & $0$ & $\frac{1}{2\sqrt{6}}(5m_K^2-3m_\pi^2)$ & $0$ & $-\frac{1}{\sqrt{6}}$ \\
   & $D^+\eta\rightarrow D^+\eta$ & $0$ & $m_\eta^2$ & $-\frac{1}{3}m_\pi^2$ & $1$ & $\frac{1}{3}$ \\
\hline
 $(0,+2)$  & $D^+\pi^+\rightarrow D^+\pi^+$ & $1$ & $m_\pi^2$ & $-m_\pi^2$ & $1$ & $1$ \\
\hline
\end{tabular}
\centering
\caption{Coefficients $C_i^{jk}$ of the LO and NLO terms of the potential for $D\phi\rightarrow D\phi$ in Eq.~(\ref{eq:potential}) in the sectors with charm $C$, strangeness $S$ and charge $Q$ in the particle basis. }
\label{tab:coeff_Qbasis}
\end{table}
\newpage
\begin{table}[H]
\begin{tabular}{c l | c c c c c}
\hline
 $(S,Q)$  &  Channel & $C_{\rm LO}^{jk}$ & $C_0^{jk}$ & $C_1^{jk}$ &  $C_{\rm 24}^{jk}$ & $C_{35}^{jk}$  \\  
\hline
 $(1,0)$   & $D^0K^0\rightarrow D^0K^0$ & $0$ & $m_K^2$ & $0$ & $1$ & $0$ \\
   & $D^0K^0\rightarrow D_s^+\pi^-$ & $1$ & $0$ & $-\frac{1}{2}(m_K^2+m_\pi^2)$ & $0$ & $1$ \\
   & $D_s^+\pi^-\rightarrow D_s^+\pi^-$ & $0$ & $m_\pi^2$ & $0$ & $1$ & $0$ \\
\hline
 $(1,+1)$   & $D_s^+\pi^0\rightarrow D_s^+\pi^0$ & $0$ & $m_\pi^2$ & $0$ & $1$ & $0$ \\
   & $D_s^+\pi^0\rightarrow D^0K^+$ & $\frac{1}{\sqrt{2}}$ & $0$ & $-\frac{1}{2\sqrt{2}}(m_K^2+m_\pi^2)$ & $0$ & $\frac{1}{\sqrt{2}}$ \\
   & $D_s^+\pi^0\rightarrow D^+K^0$ & $-\frac{1}{\sqrt{2}}$ & $0$ & $\frac{1}{2\sqrt{2}}(m_K^2+m_\pi^2)$ & $0$ & $-\frac{1}{\sqrt{2}}$ \\
   & $D_s^+\pi^0\rightarrow D_s^+\eta$ & $0$ & $0$ & $0$ & $0$ & $0$ \\
   & $D^0K^+\rightarrow D^0K^+$ & $-1$ & $m_K^2$ & $-m_K^2$ & $1$ & $1$ \\
   & $D^0K^+\rightarrow D^+K^0$ & $-1$ & $0$ & $-m_K^2$ & $0$ & $1$ \\
   & $D^0K^+\rightarrow D_s^+\eta$ & $\sqrt{\frac{3}{2}}$ & $0$ & $\frac{1}{2\sqrt{6}}(5m_K^2-3m_\pi^2)$ & $0$ & $-\frac{1}{\sqrt{6}}$ \\
   & $D^+K^0\rightarrow D^+K^0$ & $-1$ & $m_K^2$ & $-m_K^2$ & $1$ & $1$ \\
   & $D^+K^0\rightarrow D_s^+\eta$ & $\sqrt{\frac{3}{2}}$ & $0$ & $\frac{1}{2\sqrt{6}}(5m_K^2-3m_\pi^2)$ & $0$ & $-\frac{1}{\sqrt{6}}$ \\
   & $D_s^+\eta\rightarrow D_s^+\eta$ & $0$ & $m_\eta^2$ & $-\frac{4}{3}(2m_K^2-m_\pi^2)$ & $1$ & $\frac{4}{3}$ \\
\hline
 $(1,+2)$   & $D_s^+\pi^+\rightarrow D_s^+\pi^+$ & $0$  & $m_\pi^2$ & $0$ & $1$ & $0$\\
  & $D_s^+\pi^+\rightarrow D^+K^+$ & $1$ & $0$ & $-\frac{1}{2}(m_K^2+m_\pi^2)$ & $0$ & $1$ \\
  & $D^+K^+\rightarrow D^+K^+$ & $0$ & $m_K^2$ & $0$ & $1$ & $0$ \\
\hline
 $(2,+1)$  & $D_s^+K^0\rightarrow D_s^+K^0$ & $1$ & $m_K^2$ & $-m_K^2$ & $1$ & $1$ \\
\hline
 $(2,+2)$  & $D_s^+K^+\rightarrow D_s^+K^+$ & $1$ & $m_K^2$ & $-m_K^2$ & $1$ & $1$ \\
\hline
\end{tabular}
\centering
\caption{Continuation of Table~\ref{tab:coeff_Qbasis}. }
\label{tab:coeff_Qbasis2}
\end{table}

\newpage
\section{Finite-temperature modifications of light mesons}
\label{app:modpion}

In this work we have neglected the medium modifications of the light mesons and used vacuum spectral functions for them, in both the $T$-matrix calculation as well as in the $D$-meson self-energy corrections. This approximation---which should be reasonable at low temperatures---was implemented in Ref.~\cite{Montana:2020lfi}, where we based our assumption on the pion mass modifications given in Refs.~\cite{Schenk:1993ru,Toublan:1997rr}. We leave a more thorough study of the thermal modification of light mesons into our self-consistent approach for the future. In this appendix we present a validity check using a medium-modified pion mass.

To address the correction of the pion self-energy due to the thermal bath, we have applied the methodology of~\cite{Schenk:1993ru}. As opposed to our calculation for heavy mesons, the method in~\cite{Schenk:1993ru} is not self-consistent but based on the one-loop correction to the meson self-energy in the dilute limit. We have computed the real part of the pole of the pion propagator, whose self-energy is corrected by the thermal medium producing a modified dispersion relation,
\be \omega (p) \simeq \omega_p - \frac{1}{\omega_p}\int \frac{d^3q}{(2\pi)^3 2\omega_q}  f(\omega_q,T) \textrm{ Re } T_{\pi \pi} (s) \ , \ee
where $\omega_p = \sqrt{p^2+m_\pi(T=0)}$ is the vacuum dispersion relation (with $m_\pi(T=0)=138$ MeV),
$f(\omega_p,T)$ is the Bose-Einstein distribution function, $T_{\pi\pi} (s)$ is the (isospin averaged) forward amplitude of the $\pi\pi \rightarrow \pi\pi$ process and $s=(p+q)^2$ the Mandelstam variable. 

$T_{\pi\pi}(s)$ is calculated using the unitarized scattering amplitudes coming from $SU(3)$ chiral perturbation theory Lagrangian~\cite{Oller:1997ng,Oller:1998hw}. The unitarization approach used in~\cite{Oller:1997ng,Oller:1998hw} is similar (but not equal) to ours. In particular, the scattering amplitudes from~\cite{Oller:1998hw} have no corrections due to the temperature, but this is consistent with the one-loop approximation for the pion self-energy.

In this appendix we neglect the pion width---which is also generated due to temperature effects---so we can still use Dirac delta spectral functions peaked at $\omega(p)$. We define the thermal pion mass as the value $m_\pi(T) = \omega (p=0;T)$ and plot it in Fig.~\ref{fig:pionmass} up to $T=150$ MeV. At this temperature the pion mass is $m_\pi(T=150 {\rm MeV})=120$ MeV.

\begin{figure}[htp!]
\includegraphics[width=0.55\textwidth]{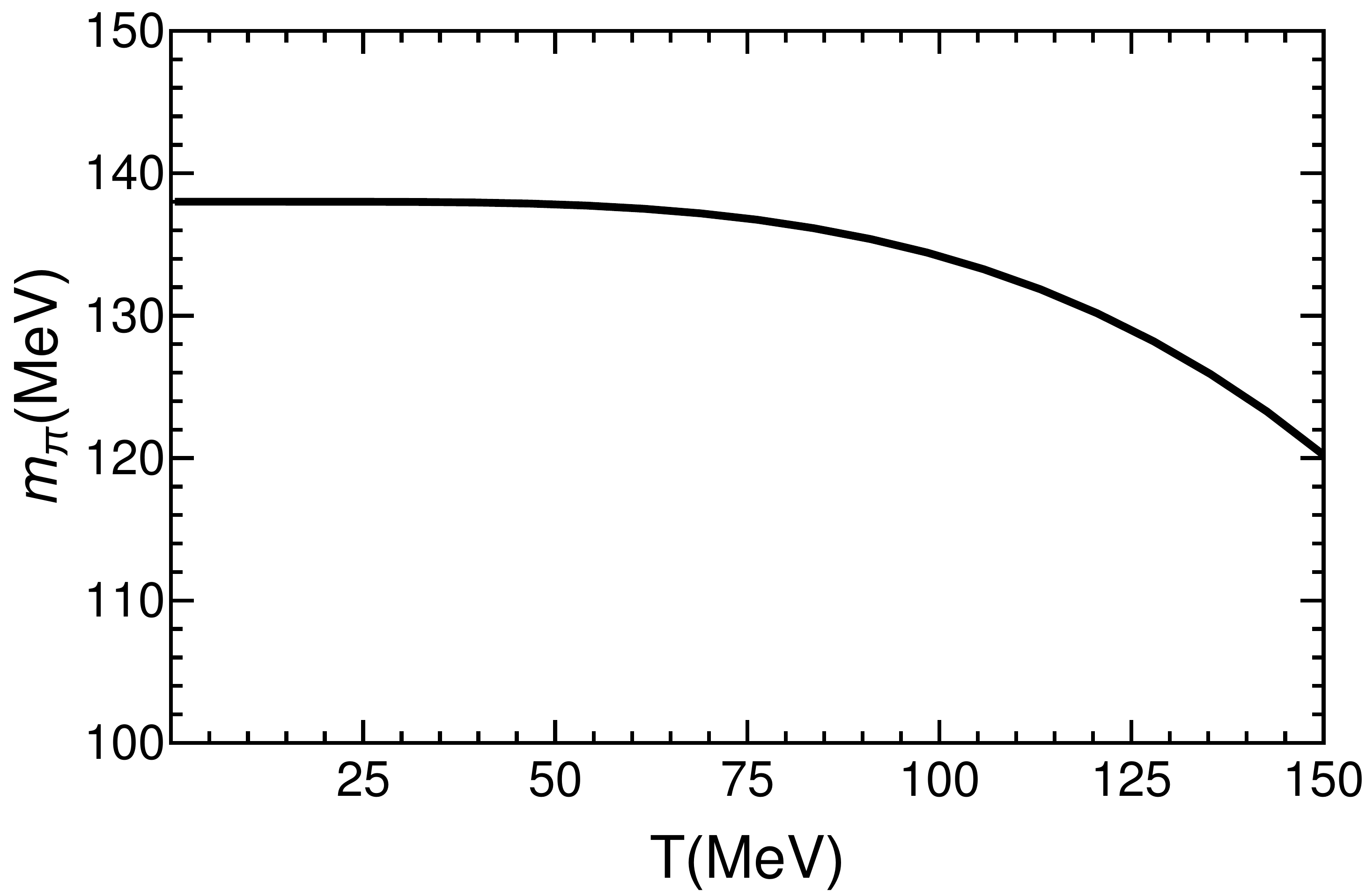}
\caption{Pion mass $m_\pi(T)=\omega(p=0;T)$ as a function of the temperature after incorporating $SU(3)$ ChPT amplitudes of Ref.~\cite{Oller:1998hw} into the one-loop pion self-energy correction of Ref.~\cite{Schenk:1993ru}.}
\label{fig:pionmass}
\end{figure}

We have run our code for the $D$-meson self-energy at $T=150$ MeV using this reduced pion mass (the pion decay will be added in a future study). We find that the mass of the ground-state $D^{(*)}$ and $D_s^{(*)}$ mesons are only slightly modified with a decrease of $\Delta m_{D^{(*)}}=4$ MeV and $\Delta m_{D_s^{(*)}}=2$ MeV, with respect to the thermal masses reported in Sec.~\ref{sec:thermalmasses}, while the widths do not change appreciably. With regards to the dynamically generated states, the lowest-lying state that corresponds to $D^*(2300)$ as well as the one for $D_1^*(2430)$ change their masses by $-2$ MeV, being the widths 20 MeV larger. As for the highest-lying resonances both change by  $-2$ MeV, with a similar change in width. For the bound states $D_{s0}^*(2317)$ and $D_{s1}^*(2460)$ the change in mass is of  $-2$ MeV, while the width increases by $1$ MeV.

In conclusion, for low temperatures $T \ll 150$ MeV it is acceptable to neglect the medium effects of the light mesons. For the largest temperature considered $T=150$ MeV, the effects of a medium-modified pion are noticeable, but still small. In the future, we plan to incorporate the medium-modified spectral functions (with both mass and decay width depending on temperature), to decide whether the widening of the pion (and also the modification of the other light mesons) can produce a significant change on the properties of heavy-flavor mesons at intermediate temperatures.

\bibliographystyle{ieeetr}
\bibliography{D-meson}
  
\end{document}